\documentclass[journal]{IEEEtran}

\ifCLASSOPTIONcompsoc
  \usepackage[nocompress]{cite}
\else
  \usepackage{cite}
\fi
\ifCLASSINFOpdf
\else
\fi
\usepackage[T1]{fontenc}
\DeclareFontFamily{T1}{calligra}{}
\DeclareFontShape{T1}{calligra}{m}{n}{<->s*[1.44]callig15}{}
\DeclareMathAlphabet\mathcalligra   {T1}{calligra} {m} {n}
\DeclareMathAlphabet\mathzapf       {T1}{pzc} {mb} {it}
\DeclareMathAlphabet\mathchorus     {T1}{qzc} {m} {n}
\DeclareMathAlphabet\mathrsfso      {U}{rsfso}{m}{n}
\usepackage{graphicx}
\usepackage{amsmath}
\usepackage{subfigure}
\usepackage{amsfonts,amsmath}
\usepackage{amssymb, nccmath}
\usepackage{amssymb, nccmath}
\usepackage{dsfont}
\usepackage{booktabs}
\usepackage{multirow}
\usepackage[utf8]{inputenc}
\usepackage{dsfont}
\usepackage{booktabs}
\usepackage{graphicx}
\usepackage{grffile}
\usepackage[linesnumbered, ruled]{algorithm2e}
\usepackage{amsthm}
\newtheorem{definition}{Definition}[]
\usepackage{grffile}
\usepackage{array} 
\usepackage{mathtools}
\usepackage{fontenc}
\usepackage{tikz}
\usepackage{lineno}
\usepackage{pgfplots}
\usetikzlibrary{patterns}
\usetikzlibrary{shapes,arrows}
\usetikzlibrary{shapes,snakes}
\usetikzlibrary{positioning}
\definecolor{mycolor1}{HTML}{F5F5DC}
\definecolor{Mycolor}{HTML}{8A2BE2}
\makeatletter
\newcommand{\removelatexerror}{\let\@latex@error\@gobble}
\def\ps@IEEEtitlepagestyle{%
	\def\@oddfoot{\mycopyrightnotice}%
	\def\@oddhead{\hbox{}\@IEEEheaderstyle\leftmark\hfil\thepage}\relax
	\def\@evenhead{\@IEEEheaderstyle\thepage\hfil\leftmark\hbox{}}\relax
	\def\@evenfoot{}%
}

\def\mycopyrightnotice{%
	\begin{minipage}{\textwidth}
		\centering \scriptsize
		This article has been accepted in IEEE Transactions on Network and Service Management Journal © 2022 IEEE. Personal use of this material is permitted. Permission from
		IEEE must be obtained for all other uses, in any current or future media, including reprinting/republishing this material for advertising or promotional purposes, creating new collective works, for resale or redistribution to servers or lists, or reuse of any copyrighted component of this work in other works. This work is freely available for survey and citation.
		
	\end{minipage}
}
\makeatother
\usepackage{subfig}
\definecolor{Mycolor}{HTML}{800000}
\renewcommand\thesection{\arabic{section}}

\renewcommand{\thesubsubsection}{\Alph{subsubsection}}

\begin{document}
%\setpagewiselinenumbers
%\modulolinenumbers[1]
%\linenumbers
%\linenumbers
%\title{A Secure and Multi-objective Virtual Machine Consolidation Framework for Cloud Datacenter}
\title{A Fault Tolerant Elastic Resource Management Framework  Towards High Availability of Cloud Services}
%\title{	OFP-RM: An Online Failure Prediction based Resource Management Framework for High Availability of Cloud Services }
%\title{COP-RM: Cloud Outage Prediction based Resource Management Framework for High Availability of Cloud Services }
%\title{COM-HA: Cloud Outage Management Framework for High Availability}
\author{Deepika~Saxena,~Ishu~Gupta,~\textit{Member, IEEE,}~Ashutosh~Kumar~Singh,~\textit{Senior Member, IEEE,}~and~Chung-Nan~Lee,~\textit{Member, IEEE}
	%and~Jane~Doe,~\IEEEmembership{Life~Fellow,~IEEE}% <-this % stops a space
	\IEEEcompsocitemizethanks{\IEEEcompsocthanksitem D. Saxena and A. K. Singh are with the Department of Computer Applications, National Institute of Technology, Kurukshetra, India. E-mail: 13deepikasaxena@gmail.com,  ashutosh@nitkkr.ac.in \\	
	 I. Gupta and C. N. Lee are with the Department of Computer Science and Engineering, National Sun Yat-Sen University, Kaohsiung, Taiwan(e-mail: ishugupta23@gmail.com, cnlee@cse.nsysu.edu.tw
		)
}}

\markboth{IEEE Transactions on Network and Service Management}%
{Shell \MakeLowercase{\textit{et al.}}: Bare Demo of IEEEtran.cls for Computer Society Journals}
\IEEEtitleabstractindextext{%
\begin{abstract}
	Cloud computing has become inevitable for every digital service which has exponentially increased its usage. {However, a tremendous surge in cloud resource demand stave off service availability resulting into outages, performance degradation, load imbalance, and excessive power-consumption.} The existing approaches mainly attempt to address the problem by using multi-cloud and running multiple replicas of a virtual machine (VM) which accounts for high operational-cost. This paper proposes a Fault Tolerant Elastic Resource Management (FT-ERM) framework that addresses aforementioned problem from a different perspective by inducing high-availability in servers and VMs.  Specifically, (1) an online failure predictor is developed to anticipate failure-prone VMs based on predicted resource contention; (2) the operational status of server is monitored with the help of power analyser, resource estimator and thermal analyser to identify any failure due to overloading and overheating of servers proactively; and (3) failure-prone VMs are assigned to proposed fault-tolerance unit composed of decision matrix and safe box to trigger VM migration and handle any outage beforehand while maintaining desired level of availability for cloud users. The proposed framework is evaluated and compared against state-of-the-arts by executing experiments using two real-world datasets. FT-ERM improved the availability of the  services up to 34.47\% and scales down VM-migration and power-consumption up to 88.6\% and 62.4\%, respectively over without FT-ERM approach.

\end{abstract}

}

\maketitle

\IEEEdisplaynontitleabstractindextext

%\renewcommand\thesubsection{\Alph{subsection}}

% \fi
\IEEEpeerreviewmaketitle

\begin{figure}[!htbp]
	\centering
	\includegraphics[width=0.7\linewidth]{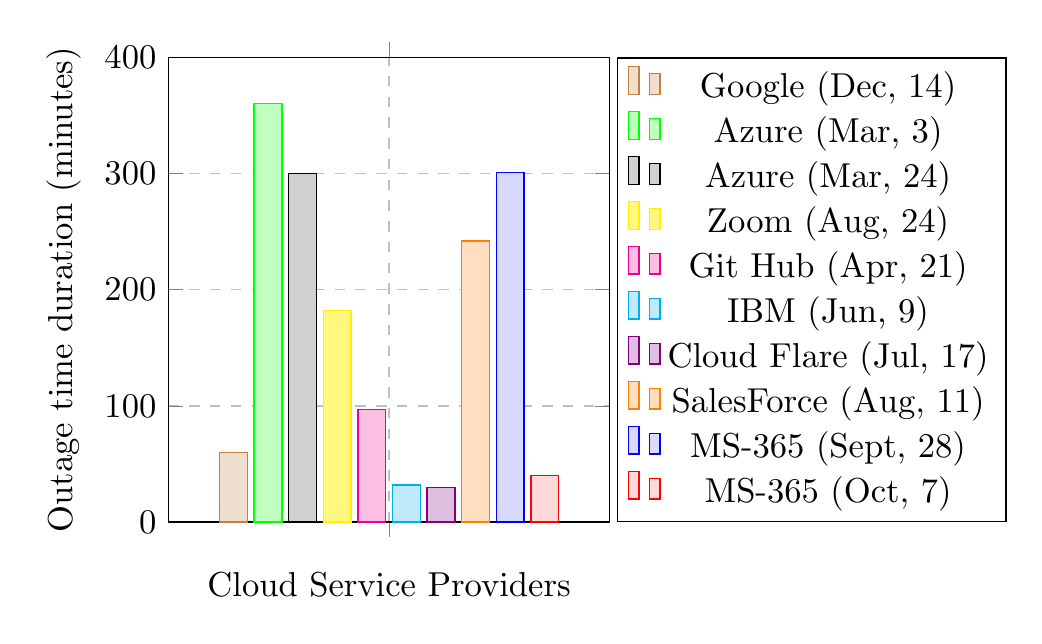}
	\caption{Biggest Cloud Outages Of 2020 }
	\label{fig:cloudoutage}
\end{figure}

%\IEEEraisesectionheading
{\section{Introduction}\label{sec:introduction}}
\IEEEPARstart{H}{ighly} available Information Technology (IT) services and maximum computing benefits at minimum capital investment is the peak concern of every cloud user. However, the delivery of a higher level of availability remains one of the biggest challenges for the Cloud Service Providers (CSPs) because of tremendously growing demand and dependability of every organization on the cloud infrastructure \cite{li2017novel}, \cite{saxena2021osc}. Though the CSP is obliged to Service Level Agreement (SLA) adherence by taking responsibility for their infrastructure and ensuring availability and safety at all ends \cite{saxena2020security}, services and performance outage occurs, stemming from a surge in resource utilization \cite{dabbagh2015energy}, \cite{saxena2021op}, \cite{singh2021quantum}, \cite{yang2020accurate}. Fig. \ref{fig:cloudoutage} reveals a recent survey of COVID-19 pandemic that witnessed several spectacular incidents of cloud outages which have dominated the titans in the area, including Zoom, Microsoft Azure, and Google Cloud Platform, Amazon Web Services, IBM Cloud etc. \cite{crn2020outage}.
 For instance, Microsoft Azure faced six hours and five hours outage on $3^{rd}$ and $24^{th}$ March, 2020, respectively because of a failure of cooling system and subsequent thermal spikes throughout the data center, hampered performance of network devices, rendered compute and storage instances inaccessible. These cloud outages could increase in frequency and severity and raises a critical challenge for CSP i.e., how to combat such outages and boost the reliance of users on cloud technology? 
\par {The aforesaid question grants a motivation for developing an elastic resource management approach to furnish High Availability (HA) and ensure reliability of cloud services. Since the availability of cloud services depends on the average availability of compute, storage, and network devices, it becomes an obligation to administer the failure governing features including mean time between failures (${MTBF}$) and mean time to repair ($MTTR$) to reduce the rate of cloud outages.} The average availability of a software or a hardware boosts with rising value of ${MTBF}$ and falls down with increasing value of $MTTR$ \cite{endo2017highly}, \cite{jammal2017evaluating}. Therefore, the key solution for improving the availability of cloud services is to maximize $MTBF$ and minimize $MTTR$ by proactive detection of failures, monitoring, analysing historical and current performance of physical machines, and concurrent handling of outages before occurrence. To accomplish the same, HA awareness feature must be embedded within the infrastructure including servers and VMs. The existing works have resolved the cloud outages by applying proactive as well as reactive approaches \cite{mukwevho2018toward}. Proactive techniques of failure handling rely on the prior knowledge of the failure of applications and VMs \cite{pinciroli2020cedule+}. The most common reasons for service outage are resource contention, over-utilization, hardware failures, inappropriate installation of softwares, or execution time exceeding the threshold value, physical machine running out of memory space, and so on \cite{jhawar2012fault}, \cite{saxena2022tellm}. The reactive techniques include checkpointing, replication, VM migration, application resubmission etc. which are triggered at the occurrence of actual failure \cite{sampaio2018comparative}. 

\subsection{Our Contributions}
\begin{figure} [!htbp]
 	\centering
 	\includegraphics[width=0.99\linewidth]{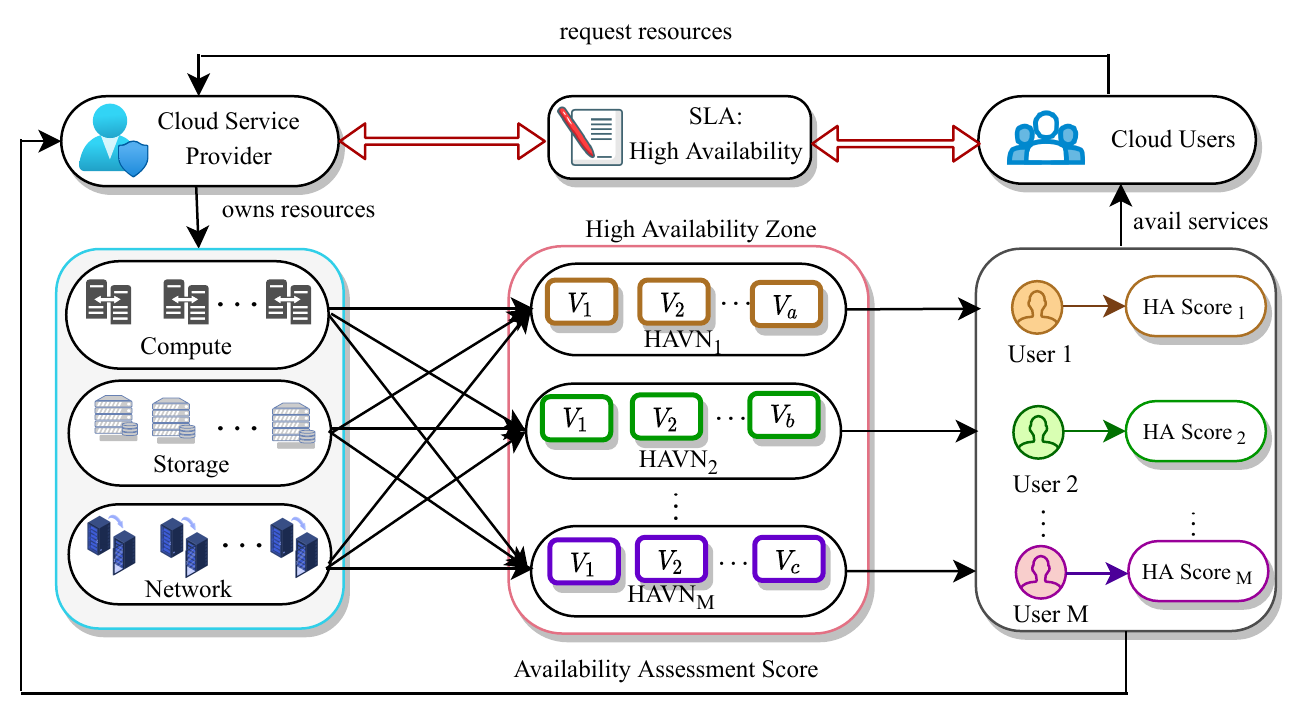}
 	\caption{Bird Eye View of Proposed Framework}
 	\label{fig:bird-eye-view}
 \end{figure}

This paper resolves cloud outage problem by proposing a novel  \textbf{F}ault \textbf{T}olerant \textbf{E}lastic \textbf{R}esource \textbf{M}anagement \textbf{(FT-ERM)} framework which embeds HA awareness in servers as well as VMs. The online monitoring features are ingrained within each server to estimate their operational status periodically based on power consumption, resource usage and thermal analysis. A dedicated High Availability Virtual Network (HAVN) is designed for each user to improve their experience of service availability. The framework employs a proposed online Multi-Input and Multi-Output Evolutionary Neural Network (MIMO-ENN) based predictor to analyse resource contention based failure of VMs proactively. The failure prone VMs are assigned to a {Fault Tolerance Unit (FTU)} for handling any outage in advance and maintaining the desired level of availability for the cloud users. The key contributions of the proposed framework include:

\begin{itemize}
\item A concept of High Availability Zone (HAZ) composed of a number of High Availability Virtual  Networks (HAVNs), is introduced to monitor and improve the HA service experience for the cloud users. 

	\item An online MIMO-ENN based Failure Prediction Unit is developed to forecast the multiple resource usage of a VM concurrently and estimate its failure status proactively. 

	\item A Failure Tolerance Unit is employed to trigger the necessary failure elimination actions and decide the safer allocation for the predicted failure prone VMs.
	 \item The performance evaluation of FT-ERM framework by using real benchmark datasets reveals that it outperforms the state-of-art approaches in terms of various performance metrics. %like Service availability, VM failure reduction, server overload prediction, resource utilization, reduction of power consumption and VM migration cost. 

\end{itemize}
A bird eye view of the proposed framework is presented in Fig. \ref{fig:bird-eye-view}, where cloud users request resources including compute, storage, and network from CSP. Both CSP and user are connected via Service Layer Agreement (SLA) which embodies HA as a major constraint. The cloud users avail services in the form of VM instances deployed on different physical machines. A HA zone comprised of several HA virtual networks \{$HAVN_1$, $HAVN_2$, ..., $HAVN_M$\}, is created to impose HA constraint on the underlined VMs owned by the users in a strict manner. HA Score is generated as a feedback to the CSP for further improvement of services. CSP intelligently manages the resources to consistently offer HA to the user and optimize the operational cost of the datacenter.      
 
\textit{Organization}: Section 2 entails a recent related work followed by a comprehensive description of the proposed framework in Section 3. Online failure prediction by developing MIMO-ENN predictor is given in Section 4. Section 5 presents failure tolerance. The performance evaluation is presented in Section 6. The conclusive remarks and future scope of the proposed work are discussed in Section 7. %Table \ref{table:notation} shows the list of symbols with their explanatory terms that have been used throughout the paper.
%\begin{table}[htbp]
%	\centering 	
%	\caption[Table caption text] {Notations}  %\cite[p.10]{refid} }
%	\label{table:notation}
	%\resizebox{8cm}{!}{
%	\resizebox{1.0\textwidth}{!}{\begin{minipage}{\textwidth}
%			\begin{tabular}{ll}
%				\hline
				%\multicolumn{2}{c}{Item} \\
				%\cline{1-2}
%				Symbols    & Explaination terms  \\
%				\hline
%				$S$, $V$, $U$   & server, virtual machine, user \\
%				$P$, $Q$, $M$   & number of servers, virtual machines, users \\
%					$\mathds{RE}$, $\mathds{PA}$, $\mathds{TA}$   & resource estimator, power analyser, temperature analyser \\	
%					$RU$, $PW$, $T$   & resource utilization, power consumption, temperature \\
%						$Z$, $\mathds{R}$   & no. of active servers, set of resources  \\
%							$\mathds{D}$, $d_i$  & set of data samples, $i^{th}$ data sample  \\
%								$l$, $h$, $o$& set of nodes in input, hidden, output layers\\
%							$\mathds{N}$, $n$ & no. of neural networks, no. of resources\\
							
%							$\delta$, $\mathds{E}_{rmse}$ & neural weight connection, root mean %squared error\\
%				$A_o$, $A_g$    & availability offered, availability guaranteed\\

%				$K$, $\mu_k$ & number of clusters, centroid of $k^{th}$ cluster\\
%				$Z_{actual}$ and $Z_{predicted}$  & actual and predicted output\\
%				$V^F$& Failure-prone VM\\
			
%				$\gamma$ & status of server\\
%				$\omega_{ik}$ & mapping of $i^{th}$ VM on $k^{th}$ server\\
				
%				$L$ & size of neural network\\

%					$C$, $M$, $BW$  & CPU, Memory, Bandwidth\\
%				\hline
%			\end{tabular}
			%}
%	\end{minipage}}
%\end{table} 
\section{Recent Related work}
%The pioneering work dealing with cloud outages includes failure prediction approaches and fault tolerance methods as follows:
%The OFP-RM Framework is compared to four state-of-the-art works which predicts failure to provide high availability: Prediction based Energy-aware Fault-tolerant Scheduling scheme (PEFS)  \cite{marahatta2020pefs}, Failure-aware energy-efficient VM consolidation (FAEE) \cite{sharma2019failure}, Hadoop Distributed Computing Clusters for Fault Prediction (HDCC) \cite{pinto2016hadoop}, and Cloud Disk Error Forecasting (CDEF) \cite{xu2018improving}. 

%Cloud outage management comprises of failure prediction and fault tolerance approaches as follows:
%\subsection{VM Failure Prediction}
Marahatta et al. \cite{marahatta2020pefs} presented a {prediction based energy-aware fault-tolerant scheduling scheme (PEFS) that applied} deep neural network based failure predictor to distinguish the upcoming workload on VMs (i.e., arriving tasks) into failure prone and non-failure prone tasks. Accordingly, two exclusive task scheduling algorithms are proposed for resource allocation where three consecutive failure-prone tasks are                                                                                                                                                                                                                                                                                                                                                                                                                                                                                                                                            replicated into three copies which are executed on different servers to resist the overlapping and redundant execution. 
Pinto et al. \cite{pinto2016hadoop} have developed {hadoop distributed computing clusters (HDCC) for fault prediction} using support vector machine (SVM) which is trained with a non-anomalous
dataset during different operation patterns like boot-up,
shutdown, idle, task allocation, resource distribution etc. to detect and classify between normal and abnormal situation. multiple additive regression trees gradient boosting (MART-GB) algorithm based cloud disk error forecasting (CDEF) approach is proposed in \cite{xu2018improving}. It ranked all disks on the basis of degree of error-proneness to allow live migration
of existing VMs and assignment of new VMs to healthy disks for improved service availability. {An early fault detection approach based on the fuzzy logic algorithm and Gaussian process (FLGP) was proposed in \cite{bui2018early} by  analysing the characteristics of failures rigorously. This approach provided an enhanced performance in terms of accuracy of failure prediction and reaction rate which consequently enhanced the availability of the cloud environment.} Adamu et al. \cite{adamu2017approach} have presented linear regression (LR) Model and 
support vector machine (SVM) i.e., {LR-SVM} with a Linear Gaussian kernel for 
predicting hardware failures in a real-time cloud environment to 
improve system availability. Lin et al. \cite{lin2018predicting} proposed a rank based node failure prediction technique {(RFPT)} by integrating LSTM model, Random Forest, and a ranking mechanism to embed intermediate results of the two models as feature inputs and ranks the nodes by their failure-proneness. 
%Li et al. \cite{li2020predicting} have utilized Artificial Intelligence for IT Operations (AIOps)AIOps (Artificial Intelligence for IT Operations),
%a recently introduced approach in DevOps, leverages data analytics and machine learning to improve the
%quality of computing platforms in a cost-effective manner. However, the successful adoption of such AIOps
%solutions requires much more than a top-performing machine learning model. Instead, AIOps solutions must
%be trustable, interpretable, maintainable, scalable, and evaluated in context.
%\subsection{Fault Tolerance approaches}
\par Sharma et al. \cite{sharma2019failure} {proposed a failure-aware energy-efficient VM consolidation (FAEE) approach which}  predicted VM failure using an exponential smoothing based forecasting technique. Accordingly, two fault tolerance methods including VM migration and VM checkpointing are triggered to handle any failure and ensure service availability. This work concluded that a significant improvement in terms of energy efficiency and reliability is achieved by considering failure characteristics of physical resources. Wang et al. \cite{wang2014festal} developed a fault-tolerant elastic scheduling algorithms (FESTAL) to provide a fault tolerant VM scheduling accompanied by virtualization technology and an appropriate VM migration scheme with battery back-up features. Zhu et al. \cite{zhu2016fault} have utilized battery back-up scheduling schemes for fault-tolerant execution of scientific workflows (FASTER) by incorporating task allocation and message transmission features that employed a backward shifting approach for use of physical resources. Sivagami et al. \cite{sivagami2019improved} have presented a dynamic fault tolerant VM migration (DFTM) algorithm for prior estimation of load on virtual links and distributing physical resources among VMs concisely. In case of VM failure, it selected supporting VM considering network topology, load distribution, and availability of physical resources. Zheng et al. \cite{zheng2011component}
  proposed a component ranking for building fault-tolerant cloud (FT-Cloud) applications that fused the system structure and application information  to identify significant components application. Thereafter, an algorithm is proposed to automatically determine an optimal fault-tolerance strategy for significant cloud components. 
     
\par Unlike existing works which have attempted to solve the challenge of VM failures by predicting single resource, the proposed approach estimates failure with enhanced proximity to the real environment by predicting multiple resources viz., CPU, memory, bandwidth concurrently. The previous works have used reactive or proactive VM migration to deal with resource contention disregarding service availability subject to SLA during predictive resource management. In contrast, the proposed FT-ERM provides a comprehensive fault tolerance solution by developing a decision matrix and determining the safe-box prior to migration of failure prone VMs so as to avoid further resource-contention and the need for VM migration frequently. Also, the prediction of VMs resource usage assists in alleviating server over-/under-load, reducing resource and power wastage. Table \ref{table:relatedwork_comparison} compares  FT-ERM with the aforementioned state-of-the-art approaches with respect to various perspectives. 

%The FT-ERM Framework is compared to four state-of-the-art works which predicts failure to provide high availability: Prediction based Energy-aware Fault-tolerant Scheduling scheme (PEFS)  \cite{marahatta2020pefs}, Failure-aware energy-efficient VM consolidation (FAEE) \cite{sharma2019failure}, Hadoop Distributed Computing Clusters for FaultPrediction (HDCC) \cite{pinto2016hadoop}, and Cloud Disk Error Forecasting (CDEF) \cite{xu2018improving}. 

\begin{table*}[!htbp]
	
	\caption{{{Comparison of FT-ERM Framework with Related Work}}  }
	\label{table:relatedwork_comparison}
	\small
	%\centering 
	%\resizebox{0.8\textwidth}{!}{\begin{minipage}{\textwidth}
	%\resizebox{9cm}{!}{
	\centering
	%	\begin{tabular}{|l||c|c||c|c|c||c|c|c|c|c|c|}
	\begin{tabular}{|p{1.8cm} |c|c||p{1.4cm}|c|c|| p{1.5cm}| p{4cm}|}
		\hline
		{{\textbf{Work}}}&\multicolumn{2}{c||}{{\textbf{Approach}}}&\multicolumn{3}{c||}{{\textbf{ Objective}}}	&\multicolumn{2}{c|}{{\textbf{Evaluation}}}
		\\ \cline{2-8}			 
		& {\textbf{Prediction}}& {\textbf{Tolerance}}& {\textbf{Failure detection}}& {\textbf{Availability}} & {\textbf{Energy}}& {\textbf{Dataset}}& {\textbf{Parameters}}\\ \hline
		
			{PEFS \cite{marahatta2020pefs}}&{$\checkmark$}&{$\checkmark$}&{$\checkmark$}&{$\times$}&{$\checkmark$}&{Eular, Internet}& {resource utilization, energy consumption, accuracy}\\ \hline
			{HDCC \cite{pinto2016hadoop}}&{$\checkmark$}& {{$\times$}} &{{$\checkmark$}} &{{$\checkmark$} }&{{$\times$} }&{{Random}} & {failure prediction accuracy,  error} \\ \hline
			{CDEF \cite{xu2018improving}} &{$\checkmark$}&  {{$\times$}} &{{$\checkmark$}} &{{$\times$}} &{{$\times$}} &{{real-world data}} &{failure prediction accuracy, error }\\ \hline
			
			{FLGP \cite{bui2018early}} &{$\checkmark$}&  {{$\times$}} &{{$\checkmark$}} &{{$\checkmark$}} &{{$\times$}} &{{Random}} &{failure prediction accuracy, prediction error} \\ \hline
			{LR-SVM \cite{adamu2017approach} }&{$\checkmark$}&  {{$\times$}} &{{$\checkmark$} }&{{$\checkmark$}} &{{$\times$} }&{{Random}} &{failure prediction accuracy, error} \\ \hline
			{RFPT \cite{lin2018predicting}} &{$\checkmark$}&  {{$\times$}} &{{$\checkmark$}} &{{$\checkmark$}} &{{$\times$}} &{{Random}} &{failure prediction accuracy, prediction error} \\ \hline
		{{FAEE \cite{sharma2019failure}}} & {$\checkmark$} &{$\checkmark$}  &{$\checkmark$} &{{$\checkmark$}}&{{$\checkmark$ }}&{Grid5000 FTA} &{resource utilization, energy consumption, accuracy} \\ \hline		
		
	{	FESTAL \cite{wang2014festal}} &{$\times$}&  {{$\checkmark$} }&{{$\times$}} &{{$\checkmark$}} &{{$\times$} }&{{Google cluster}} &{deadline, task count, execution time }\\ \hline
	{	FASTER \cite{zhu2016fault}} &{$\times$}&  {{$\checkmark$}} &{{$\times$}} &{{$\times$}} &{{$\checkmark$}} &{{Synthetic data}} &{deadline, task count, execution time}\\ \hline
		{DFTM \cite{sivagami2019improved}} &{$\times$}&  {{$\checkmark$}} &{{$\times$}} &{{$\checkmark$}} &{{$\times$}} &{{Random}} &{energy, resource usage, response time} \\ \hline
		{FTCloud \cite{zheng2011component}} &{$\times$}&  {{$\checkmark$}} &{{$\times$}} &{{$\checkmark$}} &{{$\times$}} &{{Synthetic}} &{Failure probability} \\ \hline
		{\textbf{FT-ERM}} & {$\checkmark$}&{$\checkmark$} &{ $\checkmark$} &{$\checkmark$} &{$\checkmark$} &{Google cluster, Bitbrains} &{MTTR, MTBF, availability, resource usage, power consumption, temperature, accuracy} \\ \hline

	\end{tabular}

\end{table*}	
\section{FT-ERM Framework}
\begin{figure*}[!htbp]
	\centering
	\includegraphics[width=0.7\linewidth]{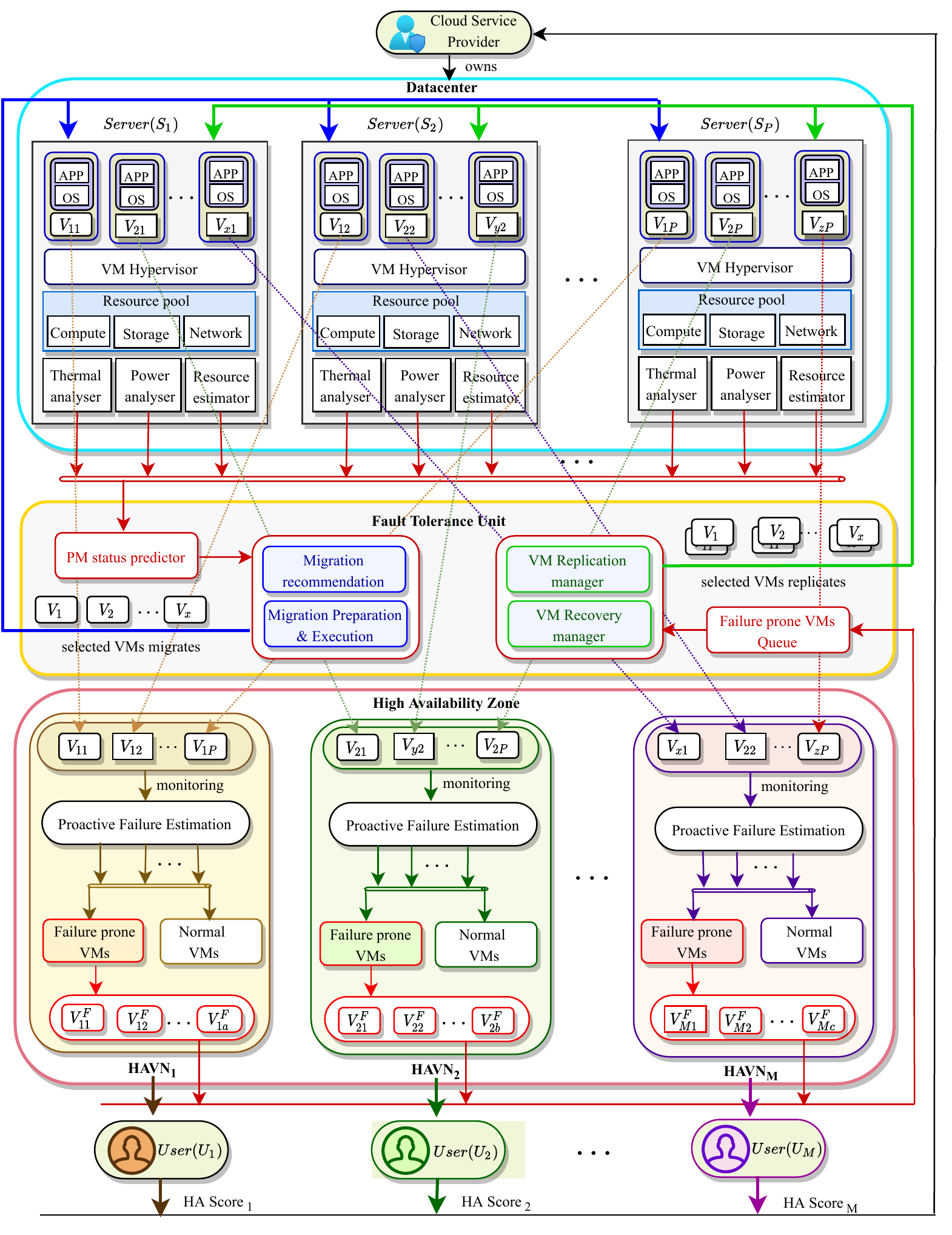}
	\caption{FT-ERM Framework}
	\label{fig:proposed-model}
\end{figure*}

{The architecture of the proposed framework 
is illustrated in Fig. \ref{fig:proposed-model} which shows the entities involved along
with the  essential
 information flow among these entities.}
Consider a CSP owns datacenter where a cluster of $P$ physical machines or servers \{$S_1$, $S_2$, ..., $S_P$\}$\subseteq \mathds{S}$ hosts $Q$ VMs 
$\mathds{V}$ of $M$ users \{$U_1$, $U_2$, ..., $U_M$\} $\subseteq \mathds{U}$. 
The VM Hypervisor layer enables deployment of VMs \{$V_{11}$, $V_{21}$, ..., $V_{x1}$\} $\subseteq \mathds{V}$ on server $S_1$ by creating an isolation layer among them. Likewise, the VMs \{$V_{12}$, $V_{22}$, ..., $V_{y2}$\} $\subseteq \mathds{V}$ and \{$V_{1P}$, $V_{2P}$, ..., $V_{zP}$\} $\subseteq \mathds{V}$ are hosted on servers $S_2$ and $S_P$, respectively. Each VM is configured with distinct guest operating system (OS) with user specified capacity of resources for the application (APP) execution. Each server comprises of a pool of compute, storage, and network resources; thermal analyser ($\mathds{TA}$), power consumption analyser ($\mathds{PA}$), and resource estimator ($\mathds{RE}$). 
\par   $\mathds{PA}$ examines the electrical power consumption ($PW$) of the server. The consumption of  power for $i^{th}$ server (i.e., $PW_i$) is computed  by applying Eq. (\ref{power1}), where ${PW_i}^{max}$, ${PW_i}^{min}$ and ${PW_i}^{idle}$ are maximum, minimum and idle state power consumption of $i^{th}$ server; $RU$ is CPU utilization of respective server. The total power consumption of entire datacenter ($PW_{dc}$) over time-interval is estimated using Eq. (\ref{power2}). 
\begin{gather}
{{PW_{i}} = 
 {[{PW_i}^{max} - {PW_i}^{min}]\times{RU} + {PW_i}^{idle}}}
\label{power1}   
\\
PW_{dc} = 
 \sum_{i=1}^{P} {{PW_i}}
\label{power2}   
\end{gather}

$\mathds{TA}$ periodically monitors the CPU temperature (${T}$) and analyses overall operational or health status ($S_j^{status}$) of the respective physical machine (PM) as stated in Eq. (\ref{tempstatus}). The term ${T}_{j}$ is current CPU temperature of $j^{th}$ PM and ${T}_{thr}$ is threshold value of server temperature, beyond which the server performance degrades.
\begin{equation}
S_j^{status} = 	\begin{cases}
Safe ($1$), & {If({T}_j < {T}_{thr}) } \\
Unsafe ($0$), & { {\textit{otherwise}} } 
\end{cases} \label{tempstatus}
\end{equation}
${T}_{j}$ is estimated based on the power consumption using the relation stated in Eq. (\ref{Temp_Power}) \cite{rezaei2019temperature}, where $\alpha$ and $\beta$ are constants, $\frac{\beta}{\alpha} = 0.130$ and according to ASHRAE \footnote{American Society of Heating, Refrigerating and Air conditioning Engineers} guidelines \cite{ashrae2011temperature}, a reference temperature for the inlet cold temperature is ${T}^{in}=20^{\circ} C$. 
\begin{equation}
	{T}_{i}=\frac{\alpha}{\beta}[\frac{PW_i-PW_i^{idle}}{PW_{i}^{max}-PW_i^{idle}}] + {T}^{in}_{i} \label{Temp_Power}
\end{equation}
$\mathds{RE}$ is employed to predict and analyse the failure status of a server because of deficient capacity of resources to fulfil the future resource requirement of VMs hosted on it. The detailed description of resource capacity utilization based failure prediction is provided in Section \ref{section:failureprediction}. The resource utilization of datacenter can be obtained using Eqs. (\ref{ru1}) and (\ref{ru2}), where $\gamma_i$ represents status of $i^{th}$ server, $Z$ is number of resources. Though in formulation, only CPU (${C}$), bandwidth ($BW$), and memory (${M}$) are considered, it is extendable to any number of resources.
\begin{gather}
{RU}_{dc}=  (\frac{\sum_{x=1}^{Z}{RU}_{x}^{{\mathds{R}}}}{|Z|\times \sum_{i=1}^{P}{\gamma_i}}) \quad \mathds{R} \in {C, M, BW}   \label{ru1}
\\
{RU}_{x}^{\mathds{R}}=\sum_{i=1}^{P}{\frac{\sum_{j=1}^{Q}{\omega_{ji} \times v_j^{\mathds{R}}}}{S_i^{\mathds{R}}}} \quad \mathds{R} \in {C, M, BW} \label{ru2}
\end{gather}
  %The proposed framework involves the basic understanding of the following definitions:
  
 The proposed framework involves HAVN, HA-Score, High Availability Zone (HAZ) which are defined as follows:
 
 \begin{definition}
	{HAVN}: A high availability virtual network defines a set of VMs \{$V^k_1$, $V^k_2$, ..., $V^k_i$\} having heterogeneous resource \{$C$, $M$, $BW$\} capacities, deployed on different physical machines by the CSP, are  assumed to be connected in a virtual network and dedicated to $k^{th}$ user $U_k$ to provide HA service.
	
\end{definition}   

\begin{definition}
	HA Score: A performance metric to quantitatively measure the quality and effective time of service availability for $k^{th}$ $HAVN$ dedicated to user $U_k$. It can be computed by applying Eqs. (\ref{avl 1}) and (\ref{avl 2}) where $A_g$ and $A_o$ are guaranteed (according to SLA terms) and offered availability, respectively. 
	\begin{gather}
	HA_{Score}=\frac{A_g -A_o}{A_g}	\times 100 \label{avl 1}
	\\
	A_o = 1-\frac{service \quad downtime}{service \quad uptime} \label{avl 2}
	\end{gather}
\end{definition}
\begin{definition}
	HAZ: A High Availability Zone is defined as a collection of virtual machines confined to provide HA to the number of users \{$U_1$, $U_2$, ..., $U_M$\} in the form of their respective HAVNs \{$HAVN_1$, $HAVN_2$, ..., $HAVN_M$\}. % where an exclusive virtual network confined to HA i.e., HAVN is designed for each user  $U_1$, $U_2$, .... $U_M$ 
\end{definition} 
 \par Consider a $HAZ$ comprising of $HAVN_1$, $HAVN_2$, ..., $HAVN_M$ dedicated to $U_1$, $U_2$, ..., $U_M$, respectively, is designed, where a set of VMs \{$V_{11}$, $V_{12}$, ..., $V_{1P}$\} of user $U_1$ and \{$V_{21}$, $V_{y2}$, ..., $V_{2P}$\} of user $U_2$ constitute $HAVN_1$ and $HAVN_2$, respectively and so on. The resource ($C$, $M$, $BW$) utilization of all the VMs belonging to each $HAVN$ are periodically monitored and a probability of failure is estimated proactively. Accordingly, the respective set of VMs is categorized into two sets including \textit{Failure prone VMs} and \textit{Normal VMs}. The Normal VMs continue to operate at the same server until it is terminated by the user. On the contrary, assignment of failure prone VMs are decided by the FTU which executes actions such as VM replication, recovery and migration to prevent the respective failure beforehand \cite{azizi2020grvmp}. 
\par {The information of predicted resource usage, estimated CPU temperature and power consumption on each server, generated by $\mathds{RE}$, $\mathds{TA}$, and $\mathds{PA}$, respectively, is passed to FTU, wherein PM status predictor estimates any server failure proactively. If PM status predictor analyses any fault due to hard disk (memory), networking device (network), and CPU (compute) failure or high temperature; the migration of all failure prone VMs is triggered from the respective PM to selected energy-efficient PMs {by following the power-efficiency concept mentioned in \cite{azizi2020grvmp}}. The energy-efficient PM signifies servers that may accomodate migrating VM such that overall power consumption of the data center get reduce.  Also, the VMs which are estimated to have a resource-contention failure, are added to queue of Failure-prone VMs which are passed to VM Replication manager to create replicas  and assign them to available servers by applying the mechanism explained in Section \ref{ftu}.  The essential constraints that must be satisfied before VM placement and migration are stated in Eq. (\ref{VMP}), where ${\mathds{R}}$ represents resources viz. CPU ($C$), memory ($M$) and bandwidth ($BW$) for assignment of VM ($V_i$) on server ($S_k$).  The users \{$U_1$, $U_2$, ..., $U_M$\} evaluate the performance of their respective \{$HAVN_1$, $HAVN_2$, ..., $HAVN_M$\} by computing HA Score using Eqs. (\ref{avl 1}) and (\ref{avl 2}). }
	\begin{gather}
	\sum_{i=1}^{Q}{V_i^{{\mathds{R}}} \times \omega_{ik} \le S_k^{{\mathds{R}}}}; \quad \forall_k \in \{1, 2, ..., P\} \label{VMP}
	\end{gather}

\section{VM Failure Prediction} \label{section:failureprediction} 
\begin{figure}[!htbp]
	\centering
	\includegraphics[width=0.95\linewidth]{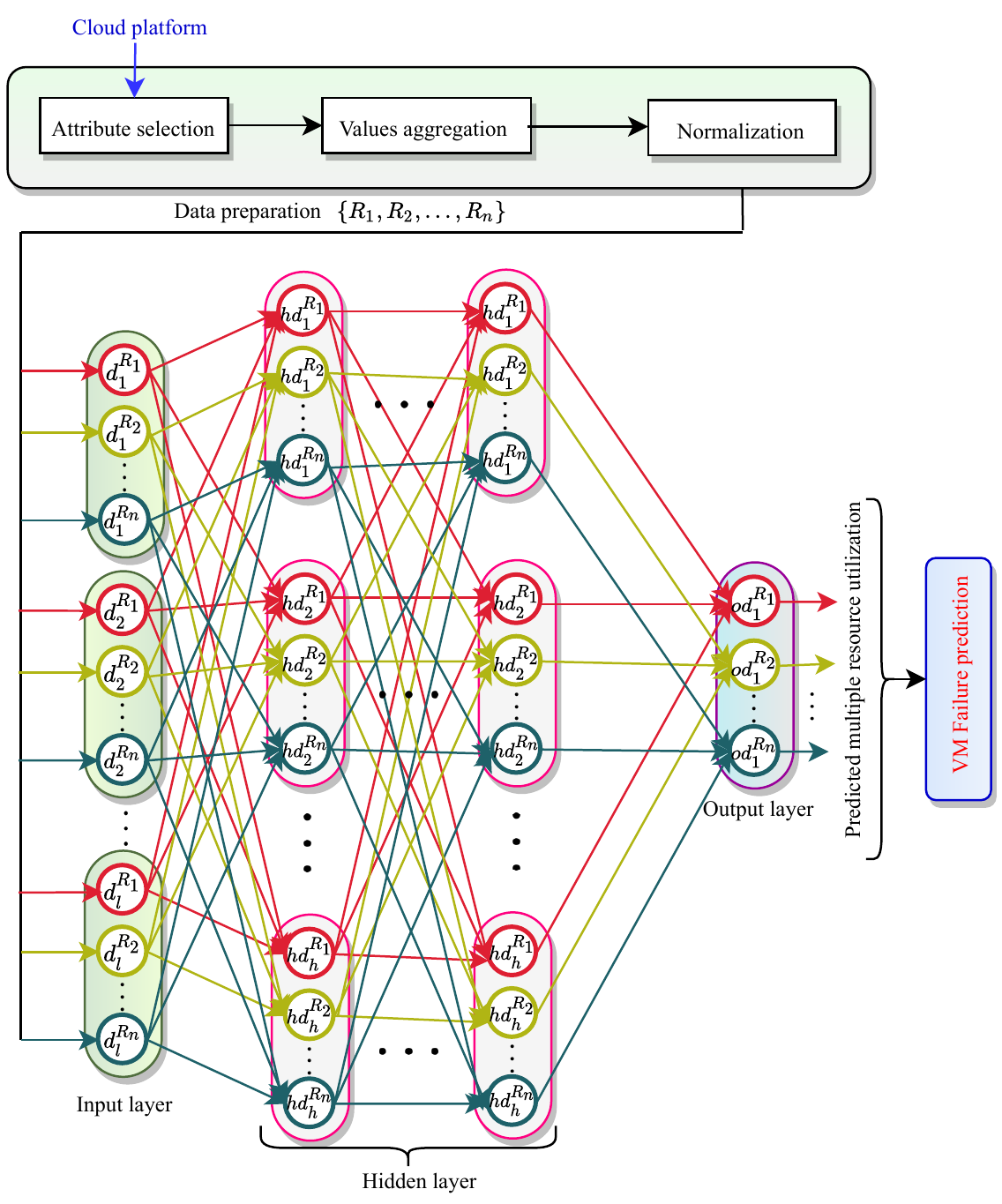}
	\caption{MIMO-ENN}
	\label{fig:failure-predictor-updated}
\end{figure}

{A \textit{Multi-Input and Multi-Output Evolutionary Neural Network} (MIMO-ENN) is developed to analyse multiple resource (CPU, memory, bandwidth, etc) usage  precisely and estimate resource capacity contention based failure respective to each VM. MIMO-ENN is a feed-forward network composed of distinct neuron network associated to a specific resource. Therefore, MIMO-ENN is a Network of multiple Networks (NoN)s, where the number of networks is  equivalent to the number of resources. The input, hidden, and output layers comprise of a set of neuron nodes instead of nodes. This set contains a distinct node respective to each intended resource. The network weight connections between two following layers link $i^{th}$ node of  $k^{th}$ set of nodes of one layer with the $i^{th}$ node of  $k^{th}$ set of nodes of consecutive layer. During the learning process, all the networks are optimized concurrently with the help of an evolutionary optimization algorithm.}  The operational flow of MIMO-ENN based failure predictor including its three key operations: \textit{data preparation}, \textit{learning}, and \textit{output generation}, is presented in Fig. \ref{fig:failure-predictor-updated}. The data is prepared from the historical capacity usage information of multiple resources \{$R_1$, $R_2$, ..., $R_n$\} $\in \mathds{R}$, gathered from the respective users' VMs deployed on different servers. The data preparation comprises three consecutive operations viz., \textit{attribute selection}, \textit{aggregation} of resource usage values per unit time, and \textit{normalization} of aggregated values in the range [0, 1]  by using Min-Max scaling equation. The vector $\mathds{D}^{{R}_i}$ is a set of normalized data values of a particular resource utilization such that $\{d^{{R}_i}_1, d^{{R}_i}_2, ..., d^{{R}_i}_l\} \in \mathds{D}^{{{R}_i}}$.
  %by applying Eq. \ref{eqn:Normalization}
%\begin{equation}
%\label{eqn:Normalization}
%\mathds{D}= \frac{ D_i- D_{min}}{D_{max}-D_{min}}
%\end{equation}
%where $D_{min}$ and $D_{max}$ are the minimum and maximum values, respectively of the input data set. 
Let the ${l+1}^{th}$ utilization of a resource depends on the previous $l$ utilization of the respective resource arranged in two dimensional input (${\mathds{D}}^{{{R}_i}}_{input}$) and output (${\mathds{D}}^{{{R}_i}}_{output}$) matrix as shown in Eq. (\ref{input}), where ${R_i}$ represents capacity utilization of $i^{th}$ resource.
\begin{equation} \label{input}
\resizebox{0.43\textwidth}{!}{$  
	%\label{eqn:7}
	{\mathds{D}}^{{{R}_i}}_{input}= 
	\left[ {\begin{array}{cccc}
		{d}^{{R_i}}_1 & {d}^{{R_i}}_2 & .... & {d}^{{R_i}}_l\\
		{d}^{{R_i}}_2 & {d}^{{R_i}}_3 & .... & {d}^{{R_i}}_{l+1} \\
		.    &    .     & .... &    .     \\
		.    &    .     & .... &    .     \\
		%   .    &    .     & .... &    .     \\
		{d}^{{R_i}}_m & {d}^{{R_i}}_{m+1} & .... &{d}^{{R_i}}_{l+m-1} \\    
		\end{array} } \right]
	%\end{bmatrix}
	% \begin{bmatrix}
	{\mathds{D}}^{{{R}_i}}_{output}= 
	\left[ {\begin{array}{c}
		{d}^{{R_i}}_{l+1} \\
		{d}^{{R_i}}_{l+2}  \\
		.      \\
		.        \\
		
		{d}^{{R_i}}_{l+m}    
		\end{array} } \right]$}
%  \end{bmatrix}
\end{equation}
%The proposed framework utilizes an ensemble approach that involves three base prediction models or experts or learners: (\textit{i}) Online Multi-resource Feed-forward Neural Network (OM-FNN) from our previous work \cite{},  (\textit{ii}) SVM and (\textit{iii}) XGBoost to estimate the future resource usage ${RU}$ of a VM. The ﬁnal outcome of the ensemble predictor is estimated by combining the outcomes of all the base learners using a voting engine. This approach improves the accuracy of the resource deficiency based failure prediction.
%The prepared data is fed into $\mathchorus{Pr}_1, \mathchorus{Pr}_2, ..., \mathchorus{Pr}_N$ to get an estimation of the upcoming $\mathchorus{W}^{\mathds{R}}$ information. Further, the predicted output of each expert $\mathchorus{z}_1, \mathchorus{z}_2, ..., \mathchorus{z}_{N}$ is passed into voting engine to estimate the ﬁnal predicted output $\mathchorus{z}_{predicted}$. This prediction will be utilized by SLD for predictive resource scaling and minimizing the resource and power wastage due to over-subscription by the users. 

\par% \textit{OM-DNN} predictor is a feed-forward evolutionary neural network which is capable of receiving input values from multiple resources and predicting output based on multiple resources, concurrently. 

{The input, multiple hidden, and output layers have $l$, $h$ and $o$ sets of nodes, respectively where the input data vector is \{\{$d_1^{{R}_1}$, $d_1^{{R}_2}$, ..., $d_1^{{R}_n}$\}, \{$d_2^{{R}_1}$, $d_2^{{R}_2}$, ..., $d_2^{{R}_n}$\}, ..., \{$d_l^{{R}_1}$, $d_l^{{R}_2}$, ..., $d_l^{{R}_n}$\}\}; and $d_j^{{R}_i}$ is the input data point given to $i^{th}$ node of $j^{th}$ set (specifies $j^{th}$ previous utilization of $i^{th}$ resource) and $n$ is the number of resources. Most recent $l$ previous resource usage information are fed periodically as an input vector for training and retraining of the MIMO-ENN predictor to forecast the future resource requirement at $(l+1)^{th}$ instance. Hence, there are $l$ sets of $n$ nodes in the input layer.} On the same lines, there are $h$ and $o$ sets of $n$ nodes at multiple hidden layers and an output layer, respectively which are represented as \{\{$\sum{hd_1}^{{R}_1}$, $\sum{hd_1}^{{R}_2}$, ..., $\sum{hd_1}^{{R}_n}$\}, \{$\sum{hd_2}^{{R}_1}$, $\sum{hd_2}^{{R}_2}$, ..., $\sum{hd_2}^{{R}_n}$\}, ..., \{$\sum{hd_h}^{{R}_1}$, $\sum{hd_h}^{{R}_2}$, ..., $\sum{hd_h}^{{R}_n}$\}\} and  \{$\sum{od_o}^{{R}_1}$, $\sum{od_o}^{{R}_2}$, ..., $\sum{od_o}^{{R}_n}$\}, respectively. The neural weight connections ($\mathchorus{W}$) between input and hidden layers are presented as $\mathchorus{W}_{jk}^{{R}_i}$, where $i^{th}$ node of $j^{th}$ set in input layer is connected to $i^{th}$ node of $k^{th}$ set in the hidden layer. The network connections between consecutive hidden layers, and the last hidden and output layers are denoted as $\mathchorus{W}_{jk}^{{R}_i}$ such that $i^{th}$ node of $j^{th}$ set in previous layer is linked to $i^{th}$ node of $k^{th}$ set in the next layer. The performance and accuracy of MIMO-ENN is evaluated by applying the Root Mean Squared Error ($\mathds{E}_{rmse}$) given in Eq. (\ref{rmse}) where $m$ is the number of training samples, $Z_{actual}$ and $Z_{predicted}$ are actual and predicted outputs, respectively. %The predicted outcome with least fitness value is generated as final predicted output. 
%\begin{equation}
\begin{gather}
\mathds{E}_{rmse}= \sqrt{\frac{1}{m}\sum_{i=1}^{m}(Z_{actual}-Z_{predicted})^2}\label{rmse} 
%\end{equation}
\end{gather}
MIMO-ENN is trained with an advance version of differential evolutionary algorithm: Dual-adaptive Differential Evolution (DaDE) which comprises five key operations viz., \textit{initialization}, \textit{evaluation}, \textit{mutation}, \textit{crossover}, and \textit{selection}. 
The consecutive steps of DaDE algorithm are as follows: (\textit{i}) $N$ number of networks, maximum generations ($Gmax$), mutation rate and crossover rate are initialized. (\textit{ii}) All the networks are evaluated on training data by using an error evaluation function (Eq. (\ref{rmse})). (\textit{iii}) Mutation selection probability $msp$ is generated to choose one of the three optional mutation schemes to be applied on each network for generation of its mutant vector. (\textit{iv}) Thereafter, heuristic crossover operation is applied for generation of new offspring. 
(\textit{v}) Fitness of each offspring vector is evaluated using Eq. (\ref{rmse}) and most optimal solution is selected to move into next generation. (\textit{vi}) The control parameters viz., crossover and mutation rates are adaptively tuned during entire evolutionary optimization process. The dual adaptation is adopted during   mutation and generation of control parameters as follows: \textit{Adaptive Mutation} Three mutation strategies namely, $ms_1$: $DE/best/1$, $ms_2$: $DE/$current-to-best$/1$ and  $ms_3$: $DE/rand/1$ are selected for TaDE algorithm.  Eqs. (\ref{eqn:mutation1}) and (\ref{eqn:mutation2}) state $ms_1$ and $ms_2$ which exploit the best individual for generation of mutant vectors, while $ms_3$ is stated in Eq. (\ref{eqn:mutation3}) that explores entire population. 	
	\begin{gather} 
	\mathchorus{MX}_i^j = \mathchorus{X}
	_{best}^j + \mu_i \times (\mathchorus{X}_{r_1}^j -\mathchorus{X}_{r_2}^j)\label{eqn:mutation1}
	\\	
	\mathchorus{MX}_i^j = \mathchorus{X}_i^j +  \mu_i \times (\mathchorus{X}_{best}^j -\mathchorus{X}_i^j) +  \mu_i \times (\mathchorus{X}_{r_1}^j -\mathchorus{X}_{r_2}^j)\label{eqn:mutation2}
	\\	
	\mathchorus{MX}_i^j = \mathchorus{X}_{r_3}^j + \mu_i \times (\mathchorus{X}_{r_1}^j -\mathchorus{X}_{r_2}^j) \label{eqn:mutation3}
	\end{gather}
	where $\mathchorus{MX}_i^j$ and $\mathchorus{X}_i^j$ depicts $i^{th}$ mutant and current vector solution of $j^{th}$ iteration, respectively. The term $\mathchorus{X}_{best}^j$ is the best solution found so far till $j^{th}$ generation and $r_1$, $r_2$ and $r_3$ are mutually distinct random numbers in the range [1, N]. The mutation scheme is decided using a random probability vector $msp$ in the range [0, 1]. Eq. (\ref{mutation_probability}) chooses mutation strategy, where $\mathchorus{P}^\ast $ depicts selected mutation scheme.
	\begin{equation}\label{mutation_probability} 
	\mathchorus{P}^\ast =
	\begin{cases}
	ms_1, & {If(0< msp_i \leq \mathchorus{P}_1 )} \\
	ms_2, & {If(\mathchorus{P}_1  < msp_i  \leq \mathchorus{P}_1 + \mathchorus{P}_2 )} \\
	ms_3, & {\text{otherwise}}  
	\end{cases}
	\end{equation}
	where $\mathchorus{P}_1$, $\mathchorus{P}_2$ and $\mathchorus{P}_3$ are the probabilities for opting the  $ms_1 $,  $ms_2 $ and $ms_3$, respectively. In reported experiments, initially $\mathchorus{P}_1$= $\mathchorus{P}_2$= 0.33, $\mathchorus{P}_3$=0.34 to allow uniform chance of selection. 
	 Thereafter, heuristic crossover is applied to mutant vector $\mathchorus{MX}_i^j$, and its current target vector $\mathchorus{X}_i^j$ for production of new offspring $\mathchorus{CX}_i^j$ i.e., $i^{th}$ solution of $j^{th}$ generation which can bring significant diversity in search space by adding promising genetic material more closer to parent with better fitness value. The fitness values of both parent chromosomes are compared to find the better one to produce a new offspring using Eq. (\ref{heuristic}), where $\mathchorus{MX}_{better}$ and $\mathchorus{MX}_i$ are parent having better fitness and other parent vector, respectively.
		\begin{gather}\label{heuristic}
		\mathchorus{CX}_i^j = \eta_{i}^j*(\mathchorus{MX}_{better}-\mathchorus{MX}_{i}) + \mathchorus{MX}_{better}  
		%h^{\gamma}_i = \xi*(h^{\beta}_i-h^{\alpha}_i) + h^{\beta}_i    
		\end{gather}
		
	Let $s_1$, $s_2$, and $s_3$ be the number of successful candidates reached into next generation for mutation strategies $ms_1$, $ms_2$, and $ms_3$, respectively. Similarly, $f_1$, $f_2$, and $f_3 $  record the number of candidates failed to reach into next generation. The probabilities of successful offspring generated by $ms_1$, $ms_2$, and $ms_3$ mutation schemes are computed as  $ms^\ast_1 $,  $ms^\ast_2 $, and  $ms^\ast_3 $ shown in Eq. (\ref{success-failure}).

	\begin{gather}
	\begin{gathered}\label{success-failure}
	%\noindent  \resizebox{1.0\hsize}{!}
	b=2(s_2s_3 + s_1s_3 + s_2s_3) + f_1(s_2 + s_3)\\ + f_2(s_1 + s_3) + f_3(s_1 + s_2) \\
	ms^\ast_1= \frac{s_1(s_2 + f_2 + s_3 + f_3 )}{b}  \\
	%\resizebox{1.0\hsize}{!}
	ms^\ast_2= \frac{s_2(s_1 + f_1 + s_3 + f_3 )}{b}  \\
	%\resizebox{1.0\hsize}{!}
	% \rho_3=\frac{sm_3(sm_1 + fm_1 + sm_2 + fm_2 )}{b} \\
	ms^\ast_3= 1- (ms^\ast_1 +ms^\ast_2 )
	\end{gathered}
	\end{gather}

Eq. (\ref{fitness}) selects population for next generation using survival of fittest concept, where $\mathchorus{W}_i^{j+1}$ is chosen candidate for the next generation, $\mathchorus{CX}_i^j $ and $\mathchorus{W}_i^j$ are the offspring generated by crossover and  the current solution, respectively. The aforementioned steps repeat until entire population converges or the near-optimal solution is achieved. 
\begin{gather}\label{fitness}
\mathchorus{W}_i^{j+1}=\begin{cases}
\mathchorus{CX}_i^j & {(fitness(\mathchorus{W}_i^j) \leq (fitness(\nu_i^j))} \\
\mathchorus{W}_i^j   & {(otherwise.)} 
\end{cases}
\end{gather}
Algorithm \ref{algo-otp-mlb} entails the operational summary of failure prediction where time-complexity can be computed as follows: 
\begin{figure}[!htbp]
	\removelatexerror
	\begin{algorithm}[H]
		\caption{Cloud Outage Prediction ()}
		\label{algo-otp-mlb}
		
		%Initialize: $List_{\mathchorus{U}}$, $List_{\mathchorus{V}}$, $List_{\mathchorus{S}}$,  $\omega$\; 
		%Generate random population of $\omega_i : i \in [1, X]$ and proceed with  $\omega_{optimal}$ that optimizes $\mathchorus{RU}$, $\mathchorus{PW}$ and $\mathfrak{R}^{score}$ as per Eq. \ref{model}, by applying $NSGA-II$ \;		
		\For {each time-interval $\{t_1, t_2\}$}{ 
			\For {each ${HAVN}_{id}: id \in [1, M]$ }{
				\For {each ${V}_{i} \in {HAVN}_{id} : i \in [1, Q]$ }{
				Predict multiple resource usage such as $V^{new}_i(\mathds{R}) \Leftarrow$ MIMO-ENN( ) where \{$R_1$, $R_2$, ..., $R_n$ \} $\in \mathds{R}$ \;
			 Update list of predicted VMs $\mathds{V}^{predicted} \Leftarrow V_i^{new} $\;
				
			\eIf{ $V_i^{new}(\mathds{R}) >$ $ V_i^{current}(\mathds{R})$ }
			{$\mathds{V}^{Failure prone} \Leftarrow V_i^{new} $\;}{$\mathds{V}^{Normal} \Leftarrow V_i^{new} $\;}	
			}
		}
	
		\For {each server ${S}_{j}: j \in [1, P]$ }{ 
			\For {each VM $V_i$ currently allocated on $S_j$}{
				Aggregate predicted resource capacity of each resource of $V^{new}_i \in$ $\mathds{V}^{predicted}$\; 
				$Tot_{c}(R_1) \Leftarrow V^{new}_i(R_1) + Tot_{c}(R_1)$, $Tot_{c}(R_2) \Leftarrow V^{new}_i(R_2) + Tot_{c}(R_2) $, ..., $Tot_{c}(R_n) \Leftarrow V^{new}_i(R_n) + Tot_{c}(R_n)$\;
		}	
			\eIf{$S_j(\mathds{R}) >$ $Tot_{c}(\mathds{R})$ || $S^{{T}}_j \le {T}_{thr}$ }{
			Update $\mathds{V}^{Failure prone}$ by removing VMs currently placed on $S_j$ \;  }{ $S_j$ will be overloaded\;
		}
		
	}
	}
	\end{algorithm}
	
\end{figure}
{The steps 1-24 execute to determine cloud outage because of server overloading or failure-prone VMs in real-time which repeat for $t$ time-intervals \{$t_1$, $t_2$\}  and generate complexity of $\mathcal{O}(t)$. The steps 2-12 differentiate failure prone and non-failure prone VMs from $Q$ VMs that belong to $M$ HAVNs. These steps produce a time-complexity of $\mathcal{O}(M) \times \mathcal{O}(Q)$, wherein step 4 calls MIMO-ENN predictor whose time complexity depends on the size of neural network ($L=(l+1)\times h + h\times h \times tot_H +h \times o $), where $tot_H$ is total number of hidden layers, number of iterations ($G$) consumed in training of MIMO-ENN, number of resources ($n$) and number
of networks (${N}$) which becomes $\mathcal{O}(nG{N}L)$. Further, the steps 13-23 estimate overloading of $P$ servers proactively by producing a complexity of $\mathcal{O}(PQ)$. The overall time-complexity for outage prediction in entire  cloud data centre is $\mathcal{O}(tnG{N}L+ tPQM)$. This algorithm runs in the background for prediction of cloud outage due to server and VM failures.} {During prediction, the training is a periodic task and can be executed in parallel on the servers equipped with enough resources, without hampering the performance of live job execution. While the prediction unit analyses and predicts future resource usage of VMs during each prediction interval. Concurrently, it has been trained and re-trained periodically between two consecutive prediction intervals with updated training data values of most recent resource usage which improves the learning and forecasting capabilities of the proposed MIMO-ENN.   }

\section{Fault Tolerance} \label{ftu}
\begin{figure*}[!htbp]
	\centering
	\includegraphics[width=0.7\linewidth]{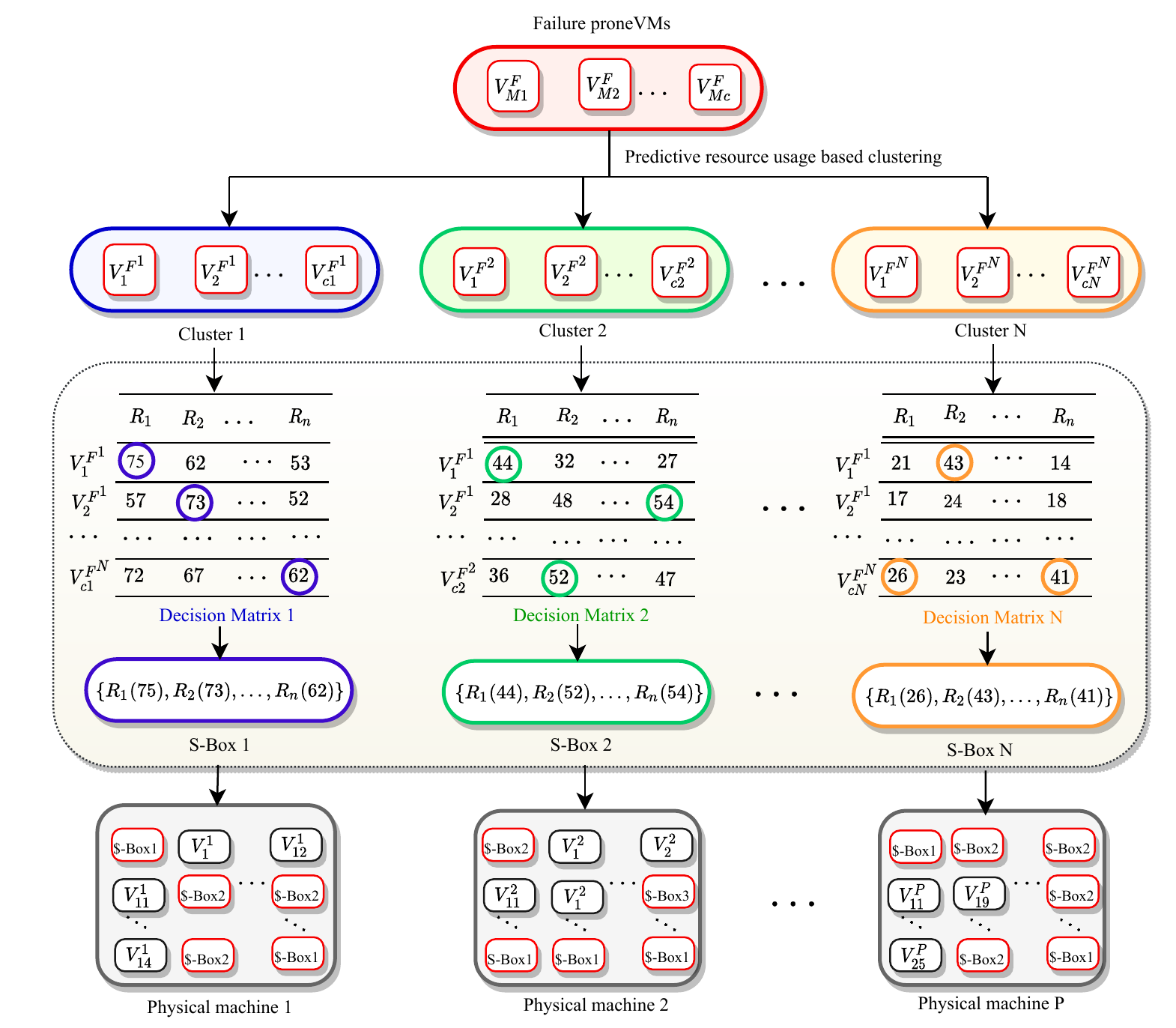}
	\caption{Fault Tolerance Unit}
	\label{fig:fault-toleranceupdated}
\end{figure*}

Let VMs \{$V_1^F$, $V_2^F$, ..., $V_{Q^\ast}^F$\} $\in \mathds{V}^F$ (where $Q^\ast$ is number of VMs) are failure prone having different capacity requirement of multiple resources beyond the currently available capacity of respective resources. Fig. \ref{fig:fault-toleranceupdated} entails the Fault Tolerance Unit, where these VMs are arranged into $K$ clusters on the basis of their predicted resources usage by applying K-means clustering algorithm. 
Elbow method \cite{saxena2020proactive} is utilized to determine effective number of clusters i.e., value of $K$ and the VMs are partitioned into $K$ pre-defined distinct non-overlapping clusters or subgroups. {K-means iterates to make inter-cluster of VMs of similar resource capacities, while keeping the clusters of VMs as different as possible. The VMs are organized into clusters by grouping the VMs according to the predicted usage capacity of resources in multi-dimension including CPU, RAM, Bandwidth} The VM is assigned to a cluster such that the sum of the squared distance between their resource requirement and centroid of the cluster is minimum by applying Eq. (\ref{eq.cluster});
\begin{equation}
\label{eq.cluster}
G=\sum_{j=1}^{Q^\ast}\sum_{k=1}^{K}{\mathds{W}_{ik}{|{V}_i^F- \mu_k|}^2}
\end{equation}
where $\mathds{W}_{ik}$ defines mapping of $i^{th}$ VM ($V_i^F$) and $\mu_k$ is centroid of $k^{th}$ cluster. Likewise, the sets of VMs \{$V_1^{F^1}$, $V_2^{F^1}$, ..., $V_{c1}^{F^{1}}$\} $\in $ $Cluster 1$, \{$V_1^{F^2}$, $V_2^{F^2}$, ..., $V_{c2}^{F^{2}}$\}$\in $ $Cluster 2$, and \{$V_1^{F^K}$, $V_2^{F^K}$, ..., $V_{cN}^{F^{K}}$\}$\in $ $Cluster K$.

A decision matrix revealing capacity requirement of $n$ resources \{$R_1$, $R_2$, ..., $R_n$\} is computed for each VM of the respective cluster to determine the physical resource requirement of the replicas of failure prone VMs. The maximum value out of each resource capacity is selected to decide the effective capacity of $n$ resources in the \textit{Safe-Box} ($S-Box$). The distinct safe boxes \{$S-Box 1$, $S-Box 2$, ..., $S-Box N$\} are estimated pertaining to each cluster. Thereafter, the replicas of VMs are assigned to the selected energy-efficient servers according to the resource capacities reserved by the S-Boxes for the respective VMs of the clusters. For instance, Decision matrix 1 reveals the resource capacity requirement of $n$ resources for each VM in Cluster 1 as shown in Fig. \ref{fig:fault-toleranceupdated}. From Decision matrix 1, the capacity requirement of $R_1$ resource is \{$75$, $57$, ..., $72$\} and it is assumed that $75$ is the highest demand for the resource $R_1$. On the same lines, the set of values \{$62$, $73$, ..., $67$\} and \{$53$, $52$, ..., $62$\} show the capacity requirement of resources $R_2$ and $R_n$, respectively for failure prone VMs \{$V_1^{F^1}$, $V_2^{F^1}$, ..., $V_{c1}^{F^{1}}$\} $\in $ $Cluster 1$. S-Box 1 is composed of highest resource capacities associated to respective resource in Decision matrix 1. The failure prone VMs of $Cluster 1$ are safely allocated by mapping ${c1}$ number of S-Boxes 1 to energy-efficient physical machines. Algorithm \ref{algo2} describes summarised steps of failure tolerance. Its time-complexity depends on number of clusters ($K$), size of decision matrix (i.e., $n \times g$), number of failure prone VMs ($Q^\ast$) and number of servers ($P$). Hence, the time complexity is $O(KPQ^\ast n \times g)$. 

\begin{figure}[!htbp]
	\removelatexerror
	\begin{algorithm}[H]
		\caption{Failure Tolerance ()}
		\label{algo2}
		{Arrange all the failure-prone VMs according to their predicted resource demand in a pool \;}
		
		{Apply K-Means Clustering algorithm to group the failure-prone VMs depending on their resource ($C$, $M$, $BW$) demand into clusters such that the clusters: \{$C_1$, $C_2$, ..., $C_K$\} $\Leftarrow \mathds{V}^{Failure prone}$ \;}

		\For {each $C_i: i \in [1, K]$}{

		Prepare a decision matrix of for cluster  $C_i$  size $n \times g$ which maps $g$  VMs to their $n$ predicted resources of VMs\;
			{Determine the maximum capacity usage of each resource separately along each column within decision matrix \;}
				{Using the maximum capacity requirement of each resource, an imaginary Safe-Box is configured \;}
					%	Determine size of $Safe-box_i \Leftarrow Max\{C_i(R_1, R_2, ..., R_n)\}$ \;

		}
	{Allocate S-boxes on available servers which maximizes resource usage and minimize power consumption while satisfying the resource capacity constraint mentioned in Eq. (\ref{VMP})}\;
	Deploy the replicas/active images of failure prone VMs as per the allocation of their respective S-boxes\;
	\end{algorithm}
	
\end{figure}      

%\subsection{Illustration}

\section{Performance Evaluation and Comparison}

\subsection{Experimental Set-up}
The simulation experiments are executed on a server machine assembled with two Intel\textsuperscript{\textregistered} Xeon\textsuperscript{\textregistered} Silver 4114 CPU with 40 core processor and 2.20 GHz clock speed. The computation machine is deployed with 64-bit Ubuntu 16.04 LTS, having main memory of 128 GB. The data center environment was set up with three different types of server and four types of VMs configuration shown in Tables \ref{table:server} and \ref{table:vm} in Python. The resource features like power consumption ($P_{max}, P_{min}$), MIPS, RAM and {storage} are taken from real server IBM \cite{IBM1999} and Dell \cite{Dell1999} configuration where $S_1$ is 'ProLiantM110G5XEON3075', $S_2$ is 'IBMX3250Xeonx3480' and $S_3$ is 'IBM3550Xeonx5675'. The VMs configuration is inspired from the VM instances of Amazon website \cite{amazon1999EC2}. 
   
\begin{table}[!htbp]
	\centering
	
	\caption[Table caption text] {Server Configuration}  %\cite[p.10]{refid} }
	\label{table:server}
		%\resizebox{0.8\textwidth}{!}{\begin{minipage}{\textwidth}
	\resizebox{9cm}{!}{
	\begin{tabular}{|lcccccc|}
		\hline
		%\multicolumn{2}{c}{Item} \\
		%\cline{1-2}
		Server&PE&MIPS&RAM(GB)&{Storage (GB)}&$PW_{max}$&$PW_{min}$/$PW_{idle}$\\
		\hline
$S_1$ 	& 2&2660&4&160&135&93.7 \\
 $S_2$	& 4&3067&8&250&113&42.3 \\
	$S_3$	& 12&3067&16&500&222&58.4 \\

		\hline
	\end{tabular}}
	%	\end{minipage}}
\end{table}

\begin{table}[!htbp]
	\centering
	
	\caption[Table caption text] {VM Configuration}  %\cite[p.10]{refid} }
	\label{table:vm}
	%	\resizebox{0.8\textwidth}{!}{\begin{minipage}{\textwidth}
	\begin{tabular}{|lcccc|}
		\hline
		%\multicolumn{2}{c}{Item} \\
		%\cline{1-2}
		VM type& PE &MIPS&RAM(GB)&Storage (GB)\\
		\hline
		$v_{small}$&1&500&0.5&40\\
		$v_{medium}$&2&1000&1&60\\
		$v_{large}$&3&1500&2&80\\
		$v_{Xlarge}$&4&2000&3&100\\

		\hline
	\end{tabular}
	%	\end{minipage}}
\end{table}
\subsection{Datasets}
The performance of proposed work is evaluated using two realworld benchmark workloads including Google Cluster Data (GCD) and Bitbrains (BB) dataset. GCD has resources CPU, memory, disk I/O  request and usage information of 672,300 jobs executed on 12,500 servers for the period of 29 days \cite{reiss2011google}. The CPU and memory utilization  percentage of VMs are obtained from the given CPU and memory usage percentage for each task in every five minutes over period of twenty-four hours. BB consists of performance metrics of fast storage 1,750 VMs from a distributed data center over period of 30 days \cite{shen2015statistical}. It contains information about CPU usage percentage, Memory usage (KB), Memory provisioned (KB), Network received and transmitted throughput (KB/s) etc. The CPU and memory usage percentage are extracted from GCD while CPU, memory, and network throughput are extracted from BB as per their availability for various experiments. The experiments are conducted for different size of data center over a period of 24 hours for GCD and 30 days for BB. 
\subsection{Simulation Configuration}
The multiple resource utilization percentage of VMs is obtained from the CPU, network, and memory usage percentage for each job in every five minutes over a period of twenty-four hours. The experiments are executed with varying sizes of datacenter such as 200, 400, 600, 800, and 1000 VMs such that the ratio of VMs:servers is 2:1, where VMs can be allocated dynamically as per demand of users with an online prediction interval of five minutes. The number of users is not mentioned in the original dataset, therefore, we have created set of user equals to 60\% of the number of VMs, requested varying number and type of VMs over time. Each user can hold VMs in the range between 0 and 10 with a constraint that at any instance, the total number of VM requests must not exceed total number of available VMs at the datacenter. {In the real-world cloud environment, the outages occur due to workload
	burst conditions, massive failure of servers, and peak of hours
	causing overloads and resource contention. Accordingly, we consider cloud outage for experiments by locating a sudden peak
	of aggregated load (or resource demand) of all VMs hosted on a server, which is
	more than the available resource capacity of the respective server, or a high CPU temperature, or a VM expecting higher resource demand in future.
	Such outages are predicted periodically, in an online service environment.} Each experiment is executed for 80 time-intervals of five minutes each to analyse the performance of proposed work dynamically, though this period can be extended as per availability of records in the dataset. Following performance metrics are evaluated: (\textit{i}) MTBF/average uptime, MTTR/average downtime, Availability, (\textit{ii}) Accuracy of Predicted vs Actual Failures,  (\textit{iii}) CPU Temperature, Resource utilization and Power consumption, 
(\textit{iv}) Number of overloads and VM  migration.

 \subsection{Results}
\begin{figure}[!htbp]
 	\centering
 	\includegraphics[width=0.7\linewidth]{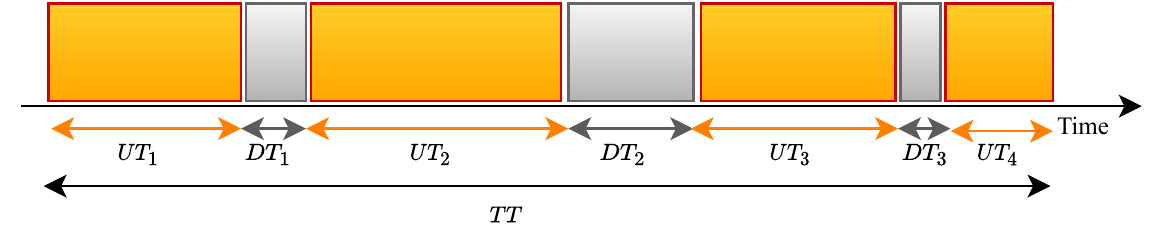}
 	\caption{MTBF and MTTR related to service uptime, outage and total time}
 	\label{fig:mttrmtbf}
 \end{figure}

 To evaluate metrics including availability, MTBF, and MTTR, the lifecycle of a hypothetical service is considered as illustrated in Fig. \ref{fig:mttrmtbf}, where the variables $DT$, $UT$, and $TT$ denote service downtime, uptime, and total time, respectively. A service can be either in downtime defined by variables $DT_1$, $DT_2$, and $DT_3$, or in uptime with variables $UT_1$, $UT_2$, $UT_3$, and $UT_4$. The term $nf$ denotes number of failures of system (i.e., 3 in Fig. \ref{fig:mttrmtbf}); the MTBF and MTTR can be computed by applying Eqs. (\ref{mtbf}) and (\ref{mttr}), respectively. These equations state MTTF and MTTR as the averages of uptime and downtime, respectively. Accordingly, the average availability can be calculated using  Eq. (\ref{availability}), where $nf$ is total number of failures, $\sum_{i=1}^{M}{UT_i}$ and $\sum_{i=1}^{M}{DT_i}$ represent total uptime and downtime, respectively experienced by $M$ users over time-interval \{$t_1$, $t_2$\}.
% The availability, MTBF,and MTTR are computed by applying Eqs. (\ref{availability})-(\ref{mttr}), where $nf$ is total number of failures, $\sum_{i=1}^{M}{UT_i}$ and $\sum_{i=1}^{M}{DT_i}$ represent total uptime and downtime respectively experienced by $M$ users over time-interval [$t_1$, $t_2$].

 \begin{equation}
 \label{mtbf}
 MTBF=\frac{\sum_{i=1}^{M}{UT_i}}{nf}
 \end{equation}
 \begin{equation}
 \label{mttr}
 MTTR= \frac{\sum_{i=1}^{M}{DT_i}}{nf}
 \end{equation}
  \begin{equation}
 \label{availability}
 A_{avg} = \frac{MTBF}{MTBF+MTTR} 
 \end{equation} 
Table \ref{table:performanceGCD} and Table \ref{table:performanceBB} report the performance metrics: \textit{MTTR}, \textit{MTBF}, average availability ($A_{avg}$), accuracy of failure prediction (\textit{$Acc^{P}$}), number of predicted failures (\textit{$Fail^{P}$}), and number of live VM migration ($ Mig\texttt{\#}$) achieved for GCD and BB workloads for varying size of datacenter (200 VMs to 1000 VMs) over period of 400 minutes. 
 \begin{table}[!htbp]
 	
 	\caption[Table caption text] {Performance metrics for GCD workloads}  
 	\label{table:performanceGCD}
 	\small
 	%\centering 
 	%\resizebox{0.8\textwidth}{!}{\begin{minipage}{\textwidth}
 	\resizebox{9cm}{!}{
 		\centering
 		\begin{tabular}{|l|c|c|c|c|c|c|c|}
 			\hline
 			
 			 VM\texttt{\#}&$T(min.)$& MTTR &MTBF &$A_{avg}$$^a$ &$Acc^{P}$$^b$ &$Fail^{P}$$^c$ &$ Mig\texttt{\#}$$^d$  \\ \hline \hline			
 			\multirow{5}{*}{200}
 			&100& 1.47&2757.14&99.94&95.5&75 &7 \\ \cline{2-8}
 			&200&1.68& 2400.00& 99.93&95.0&86 &8\\ \cline{2-8}
 			&300&1.47&2757.14&99.94&95.5& 72&7 \\ \cline{2-8}
 			&400&1.26&3233.33&99.96&99.3&96 &6\\ \hline \hline
 			
 			\multirow{5}{*}{400}
 			&100&4.41&1804.76&99.76&94.8&119 &21 \\ \cline{2-8}
 			&200&3.57&2252.94&99.84&95.8&395 &17 \\ \cline{2-8}
 			&300&3.78&2122.22&99.83&95.5&262 &18 \\ \cline{2-8}
 			&400&3.15&2566.67&99.88&96.3&140 &15 \\ \hline \hline
 			
 			\multirow{5}{*}{600}
 			&100&4.83&2508.70&99.81&96.1&583&23 \\ \cline{2-8}
 			&200&3.99& 3057.90&99.90&96.8&411 &19 \\ \cline{2-8}
 			&300&3.99&3057.90&99.90&96.8&544&19 \\ \cline{2-8}
 			&400&5.88&2042.86&99.71&95.3&536&28 \\ \hline \hline
 			\multirow{5}{*}{800}
 			&100&6.09&2658.62&99.25&96.38 &733&29 \\ \cline{2-8}
 			&200&4.41&3709.52&99.88&97.38 &725&21\\ \cline{2-8}
 			&300&5.25&3100.00& 99.83 &96.88 &696&25 \\ \cline{2-8}
 			&400&3.78&4344.44& 99.91&97.75 &713&18 \\ \hline \hline
 			\multirow{5}{*}{1000}
 			&100&6.93&3233.33&99.79&96.7 &906&33 \\ \cline{2-8}
 			&200&7.77&2602.70& 99.70&96.3&607&37\\ \cline{2-8}
 			&300&7.14&2841.18& 99.75&96.6&999&34 \\ \cline{2-8}
 			&400&5.88&3471.43&99.83&97.2 &972&28 \\ \hline 
 	\end{tabular}}
 \footnotesize{\tiny{{$^a$ Av.avg.:Availability average, $^b$ $Acc^{P}$:Failure prediction accuracy, $^c$ $Fail^{P}$: Number of predicted failures, $^d$ $ Mig\texttt{\#}$:  Number of VM migrations}}}
 \end{table}
 The prediction accuracy of multiple resources using MIMO-ENN predictor governs the performance of all other metrics. The average of failure prediction accuracy varies from 86\% to 99.3\% and 88\% to 98\% for GCD and BB workloads, respectively. The $Fail^{P}$ and $ Mig\texttt{\#}$ vary directly but non-uniformly depending upon the size of datacenter and errors in failure prediction. {The reason behind the obtained values is the periodic training and re-training of the proposed MIMO predictor  using Dual adaptive Differential Evolution optimization algorithm which enables learning of most optimal neural weights by exploring and exploiting the population of networks resulting into high prediction accuracy for multiple resources. The accurate estimation of resource contention-based failures and thereby determining suitable safe-box leads to effective scaling of VMs to mitigate the impact of VM failures proactively.}
\begin{table}[!htbp]
 	
 	\caption[Table caption text] {Performance metrics for Bitbrains workloads}  
 	\label{table:performanceBB}
 	\small
 	%\centering 
 	%\resizebox{0.8\textwidth}{!}{\begin{minipage}{\textwidth}
 	\resizebox{9cm}{!}{
 		\centering
 		\begin{tabular}{|l|c|c|c|c|c|c|c|}
 			\hline
 		
 			VM\texttt{\#}&$T(min.)$& MTTR &MTBF &$A_{avg}$&$Acc^{P}$&$Fail^{P}$&$ Mig\texttt{\#}$ \\ \hline \hline			
 			\multirow{5}{*}{200}
 			&100& 0.42&9900.00&99.99&99.89&167 &2\\ \cline{2-8}
 			&200&0.84& 4900.00& 99.98&99.15&172 &4\\ \cline{2-8}
 			&300&0.42&9900.00&99.99&99.54& 161&2 \\ \cline{2-8}
 			&400&0.63&6566.67&99.98&99.46&159 &3 \\ \hline \hline
 			
 			\multirow{5}{*}{400}
 			&100&0.42&19900.00&99.99&99.99&389 &2 \\ \cline{2-8}
 			&200&0.42&19900.00&99.99&99.60&334 &2 \\ \cline{2-8}
 			&300&0.84&9900.00&99.99&99.17&400 &4 \\ \cline{2-8}
 			&400&1.26&6566.67&99.98&98.91&370 &6 \\ \hline \hline
 			
 			\multirow{5}{*}{600}
 			&100&2.31&5354.55&99.95&98.16&514 &11 \\ \cline{2-8}
 			&200&2.73& 4515.39&99.88&97.16&511 &13 \\ \cline{2-8}
 			&300&2.73&4515.39&99.88&97.83&589&13 \\ \cline{2-8}
 			&400&3.57&3429.41&99.90&97.16&576&17 \\ \hline \hline
 			\multirow{5}{*}{800}
 			&100&3.36&4900.00&99.93&97.70 &633&16 \\ \cline{2-8}
 			&200&3.57&4605.88&99.94&98.00 &753&17\\ \cline{2-8}
 			&300&3.15&5233.33& 99.94 &98.13 &596&15 \\ \cline{2-8}
 			&400&3.99&4110.53& 99.93&97.63 &713&19 \\ \hline \hline
 				\multirow{5}{*}{1000}
 			&100&4.62&4445.45&99.89&97.80 &993&22 \\ \cline{2-8}
 			&200&2.94&7042.86& 99.95&98.60&895&14 \\ \cline{2-8}
 			&300&3.99&5163.16& 99.92&98.10&999&19 \\ \cline{2-8}
 			&400&2.73&7592.31&99.96&98.70 &977&13 \\ \hline 
 	\end{tabular}}
 \end{table}
Furthermore, to be observed that the obtained values of both MTBF and MTTR depends on the number of failures ($nf$) as depicted in Eqs. (\ref{mtbf}) and (\ref{mttr}). The values of $UT$ are obtained by computing product of {number of successfully deployed VMs} and {time-interval} over period \{$t_1$, $t_2$\}. The MTTR value associated to a VM is 0.21 minutes which is utilized from \cite{araujo2014availability}, \cite{santos2017analyzing}. Accordingly, the values of MTTR are computed for different number of VM migrations which varies with the number of unpredicted failures. The resultant availability values are computed by applying Eq. (\ref{availability}) and utilizing MTBF and MTTR values recorded during respective time-interval \{$t_1$, $t_2$\}. The availability for the two workloads is above 99\% for all the scenarios and each time-interval. \par
\begin{figure*}[!htbp]
	
	\centering
	\subfigure[Resource utilization ]{%\includegraphics[width=.32\textwidth]{Figures/RU}
	\resizebox{0.26\textwidth}{!}{\begin{tikzpicture}
\begin{axis}[
width=0.55\textwidth,
height=0.45\textwidth,
%axis lines=left,
%axis on top=true,
xmin=200,
xmax=1000,
xlabel={Size of Data center},
xticklabel style={/pgf/number format/1000 sep=},
ymin=59,
ymax=64,
ylabel={Resource utilization (\%)},
legend style={at={(0.5,0.98)},
	anchor=north,legend columns=2},
ymajorgrids=true,
xmajorgrids=true,
grid style=dashed,
]
\addplot+[color=Mycolor,mark=oplus,thick] coordinates {
	(200,60.46)(400, 60.84)(600, 62.25)(800,61.65)(1000,62.03)};\addlegendentry{GCD}
\addplot[color=blue, mark=square,thick] coordinates { 
	(200,61.68)(400, 62.23)(600, 61.92)(800, 61.97)(1000,61.81)  };\addlegendentry{BB}
  \end{axis}
\end{tikzpicture}}
}\hfill
	\subfigure[Average Temperature ]{\resizebox{0.26\textwidth}{!}{\begin{tikzpicture}
			\begin{axis}[
			width=0.55\textwidth,
			height=0.45\textwidth,
			%axis lines=left,
			%axis on top=true,
			xmin=0,
			xmax=1000,
			xlabel={Size of Data center},
			xticklabel style={/pgf/number format/1000 sep=},
			ymin=20,
			ymax=24,
			ylabel={Avg. Temperature($^{\circ} C$)},
			legend style={at={(0.5,0.98)},
				anchor=north,legend columns=2},
			ymajorgrids=true,
			xmajorgrids=true,
			grid style=dashed,
			]
			\addplot+[color=Mycolor,mark=oplus,thick] coordinates {
				(0,20) (200,22.419)(400, 22.37)(600, 22.58)(800,22.4)(1000,22.846)};\addlegendentry{GCD}
			\addplot[color=blue, mark=square,thick] coordinates { (0,20)(200,22.05)(400, 22.19)(600, 22.2348)(800,22.054)(1000,22.346)
			};\addlegendentry{BB}
			\end{axis}
			\end{tikzpicture}}}\hfill
	\subfigure[Average Power consumption ]{\resizebox{0.26\textwidth}{!}{\begin{tikzpicture}
			\begin{axis}[
			width=0.55\textwidth,
			height=0.45\textwidth,
			%axis lines=left,
			%axis on top=true,
			xmin=0,
			xmax=1000,
			xlabel={Size of Data center},
			xticklabel style={/pgf/number format/1000 sep=},
			ymin=0,
			ymax=50,
			ylabel={Power consumption (KW)},
			legend style={at={(0.5,0.98)},
				anchor=north,legend columns=2},
			ymajorgrids=true,
			xmajorgrids=true,
			grid style=dashed,
			]
			\addplot+[color=Mycolor,mark=oplus,thick] coordinates {
				(0,0) (200,7.72)(400, 17.7148)(600, 27.052)(800, 32.522)(1000,37.512)};\addlegendentry{GCD}
			\addplot[color=blue, mark=square,thick] coordinates { 
				(0,0) (200,5.09)(400, 10.0583)(600, 15.088)(800, 20.12361)(1000,25.16980)  };\addlegendentry{BB}
			\end{axis}
			\end{tikzpicture}}}

	\caption{Performance metrics for GCD and BB VM traces }
	\label{fig:ofp-rm_performancemetrics}
	
\end{figure*}
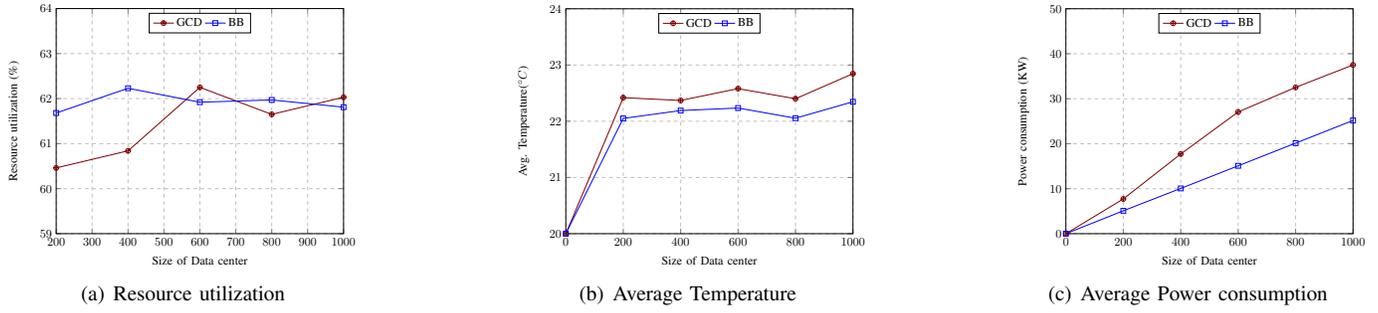

 Fig. \ref{fig:ofp-rm_performancemetrics} shows the average resource utilization, temperature and power consumption for different size of the datacenter for both workloads. The resource utilization varies from $59.3\%$ to $64.5\%$ and $61.2\%$ to $63.7\%$ for GCD and BB workloads, respectively. The average temperature varies from $22.5^{\circ} C$ to $23.6^{\circ} C$ and $24.5^{\circ} C$ to $26.6^{\circ} C$ for GCD and BB workloads, respectively. {The achieved values of resource utilization and temperature  are independent of size of data center and varies slightly according to the adopted strategy for resource distribution and VM allocation.} The power consumption increases with increasing size of datacenter depending upon the number of active servers during a particular time-interval \{$t_1$, $t_2$\}. {The resultant performance values for both the workloads vary according the real-world VMs CPU and memory usage values percentage of both the traces. The power consumption scales up with size of the data center.}

\subsection{Comparison}

\subsubsection{Failure Prediction}
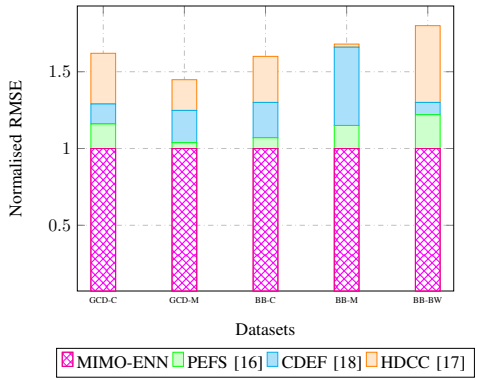
\begin{figure}
	\centering
	\resizebox{0.35\textwidth}{!}{ \begin{tikzpicture}
		\begin{axis}[
		width=0.5\textwidth,
		height=0.4\textwidth,
		%axis lines=left,
		%axis on top=true,
		ymin=0.2,
		ymax=1.8,
		ybar stacked,
		ymajorgrids=true,
		xmajorgrids=true,
		grid style=dashdotted,
		legend style={at={(0.5,-0.20)},
			anchor=north,legend columns=4},
		xtick={1,2,3,4,5},
		ylabel={Avg. active servers (\%)},
		xticklabels={\tiny{GCD-C},\tiny{GCD-M},\tiny{BB-C},\tiny{BB-M} ,\tiny{BB-BW}},
		bar width=14pt,
		%nodes near coords,
		enlargelimits=0.08,
		ylabel={Normalised RMSE},
		xlabel={Datasets},
		%xticklabel style={/pgf/number format/1000 sep=},
		%symbolic x coords={100, 200, 400, 600,		800, 1000},
		xtick=data,
		%x tick label style={rotate=0},
		]
		
		\addplot+[ybar,fill=white!20!,thin, pattern color=magenta, draw=magenta, postaction={
			pattern=crosshatch, thin}] plot coordinates {(1,1) (2,1) (3,1) (4,1) (5,1)};\addlegendentry{MIMO-ENN}
		\addplot+[ybar,fill=black, draw=green, fill=green!20!, thin] plot coordinates {(1, 0.16) (2, 0.038) (3, 0.07) (4, 0.15) (5, 0.22) };\addlegendentry{PEFS \cite{marahatta2020pefs}}
		\addplot+[ybar,fill=cyan!30!, draw=cyan, thin] plot coordinates {(1,0.13) (2,0.21) (3,0.23) (4,0.51) (5,0.08) };\addlegendentry{CDEF \cite{xu2018improving}}
		\addplot+[ybar, fill=orange!20!, draw=orange, thin] plot coordinates {(1,0.33) (2,0.20) (3,0.3) (4,0.02) (5,0.5) };\addlegendentry{HDCC \cite{pinto2016hadoop}}
		%\addplot+[ybar, draw=magenta, fill=red!20!, thin] plot coordinates {(1,0) (2,0) (3,0) (4,0) (5,0.55) };
		
		\end{axis}
		\end{tikzpicture}}
	\caption{Normalized RMSE}
	\label{fig:rmse}
\end{figure}

Figs. \ref{fig:accuracy_GCD} and \ref{fig:accuracy_BB} compare the   failure prediction accuracy of proposed MIMO-ENN of FT-ERM with the three relative methods. On average, the prediction accuracy of proposed MIMO-ENN approach varies from 95\% to 99.8\%, while the average accuracy of PEFS \cite{marahatta2020pefs}, CDEF \cite{xu2018improving}, and HDCC \cite{pinto2016hadoop} ranges from 86\% to 94\%, 81\% to 92\%, and 76\% to 84\%, respectively. The normalized RMSE values of MIMO-ENN are compared to existing methods in Fig. \ref{fig:rmse} using multiple resource information of GCD and BB workloads. The value of normalised error decreases in the order: HDCC $\leq$ CDEF $\leq$ PEFS $\leq$ FT-ERM. {The prediction accuracy of FT-ERM is highest among comparative approaches because of multi-dimensional learning process by utilizing evolutionary optimization approach for optimization of MIMO-ENN, whereas existing works HDCC, CDEF, and PEFS have used SVM, MART-GB and Deep NN which operates through conventional optimization approach based on single solution applying gradient descent method.} 
\begin{figure}[!htbp]
	
	\centering
	\subfigure[GCD-CPU ]{\resizebox{0.24\textwidth}{!}{\begin{tikzpicture}
			\begin{axis}[
			width=0.55\textwidth,
			height=0.45\textwidth,
			%axis lines=left,
			%axis on top=true,
			xmin=200,
			xmax=1000,
			xlabel={Size of Data center},
			xticklabel style={/pgf/number format/1000 sep=},
			ymin=60,
			ymax=100,
			ylabel={GCD-CPU Accuracy (\%)},
			legend style={at={(0.5,0.3)},
				anchor=north,legend columns=2},
			ymajorgrids=true,
			xmajorgrids=true,
			grid style=dashed,
			]
			\addplot+[color=magenta,mark=oplus,thick] coordinates {
				(0,0) (200,95.4)(400, 95.3)(600, 89.6)(800, 86.5)(1000,87.8)};\addlegendentry{MIMO-ENN}
			\addplot[color=green, mark=square,thick] coordinates { 
				(0,0) (200,88.5)(400, 86.1)(600,85.1 )(800, 84.361)(1000,83.980)  };\addlegendentry{PEFS \cite{marahatta2020pefs}}
			\addplot[color=cyan, mark=triangle,thick] coordinates { 
				(0,0) (200,88.5)(400, 86.9)(600,88.7 )(800, 87.361)(1000,85.80)  };\addlegendentry{CDEF \cite{xu2018improving}}
			\addplot[color=orange, mark=o,thick] coordinates { 
				(0,0) (200,80.5)(400, 79.1)(600,81.1 )(800, 80.361)(1000,81.80)  };\addlegendentry{HDCC \cite{pinto2016hadoop}}
			\end{axis}
			\end{tikzpicture}
	
}}\hfill
	\subfigure[GCD-Memory ]{\resizebox{0.24\textwidth}{!}{\begin{tikzpicture}
			\begin{axis}[
			width=0.55\textwidth,
			height=0.45\textwidth,
			%axis lines=left,
			%axis on top=true,
			xmin=200,
			xmax=1000,
			xlabel={Size of Data center},
			xticklabel style={/pgf/number format/1000 sep=},
			ymin=60,
			ymax=100,
			ylabel={GCD-Memory Accuracy (\%)},
			legend style={at={(0.5,0.3)},
				anchor=north,legend columns=2},
			ymajorgrids=true,
			xmajorgrids=true,
			grid style=dashed,
			]
			\addplot+[color=magenta,mark=oplus,thick] coordinates {
				(0,0) (200,94.4)(400, 96.3)(600, 90.6)(800, 88.5)(1000,88.8)};\addlegendentry{MIMO-ENN}
			\addplot[color=green, mark=square,thick] coordinates { 
				(0,0) (200,84.5)(400, 86.1)(600,84.1 )(800, 86.361)(1000,86.980)  };\addlegendentry{PEFS \cite{marahatta2020pefs}}
			\addplot[color=cyan, mark=triangle,thick] coordinates { 
				(0,0) (200,87.5)(400, 87.9)(600,89.7 )(800, 85.361)(1000,84.80)  };\addlegendentry{CDEF \cite{xu2018improving}}
			\addplot[color=orange, mark=o,thick] coordinates { 
				(0,0) (200,80.5)(400, 80.6)(600,80.1 )(800, 78.361)(1000,79.80)  };\addlegendentry{HDCC \cite{pinto2016hadoop}}
			\end{axis}
			\end{tikzpicture}
			
	}}\hfill

	\caption{Failure prediction accuracy for GCD workload }
	\label{fig:accuracy_GCD}
	
\end{figure}
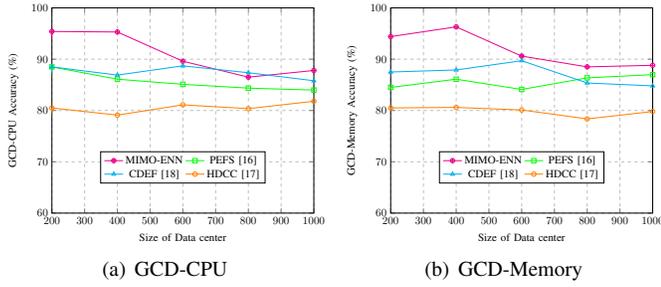 
\begin{figure}[!htbp]
	
	\centering
	\subfigure[BB-CPU ]{\resizebox{0.24\textwidth}{!}{ \begin{tikzpicture}
			\begin{axis}[
			width=0.55\textwidth,
			height=0.45\textwidth,
			%axis lines=left,
			%axis on top=true,
			xmin=200,
			xmax=1000,
			xlabel={Size of Data center},
			xticklabel style={/pgf/number format/1000 sep=},
			ymin=60,
			ymax=100,
			ylabel={BB-CPU Accuracy (\%)},
			legend style={at={(0.5,0.3)},
				anchor=north,legend columns=2},
			ymajorgrids=true,
			xmajorgrids=true,
			grid style=dashed,
			]
			\addplot+[color=magenta,mark=oplus,thick] coordinates {
				(0,0) (200,96.4)(400, 92.3)(600, 88.6)(800, 86.5)(1000,89.8)};\addlegendentry{MIMO-ENN}
			\addplot[color=green, mark=square,thick] coordinates { 
				(0,0) (200,90.5)(400, 90.1)(600,88.1 )(800, 85.361)(1000,85.980)  };\addlegendentry{PEFS \cite{marahatta2020pefs}}
			\addplot[color=cyan, mark=triangle,thick] coordinates { 
				(0,0) (200,81.5)(400, 80.9)(600,84.1 )(800, 83.361)(1000,82.980)  };\addlegendentry{CDEF \cite{xu2018improving}}
			\addplot[color=orange, mark=o,thick] coordinates { 
				(0,0) (200,75.5)(400, 80.1)(600,79.1 )(800, 77.361)(1000,81.980)  };\addlegendentry{HDCC \cite{pinto2016hadoop}}
			\end{axis}
			\end{tikzpicture}
	}}\hfill
	\subfigure[BB-Memory ]{\resizebox{0.24\textwidth}{!}{\begin{tikzpicture}
			\begin{axis}[
			width=0.55\textwidth,
			height=0.45\textwidth,
			%axis lines=left,
			%axis on top=true,
			xmin=200,
			xmax=1000,
			xlabel={Size of Data center},
			xticklabel style={/pgf/number format/1000 sep=},
			ymin=60,
			ymax=100,
			ylabel={BB-Memory Accuracy (\%)},
			legend style={at={(0.5,0.3)},
				anchor=north,legend columns=2},
			ymajorgrids=true,
			xmajorgrids=true,
			grid style=dashed,
			]
			\addplot+[color=magenta,mark=oplus,thick] coordinates {
				(0,0) (200,93.4)(400, 90.3)(600, 90.6)(800, 90.5)(1000,91.8)};\addlegendentry{MIMO-ENN}
			\addplot[color=green, mark=square,thick] coordinates { 
				(0,0) (200,87.5)(400, 88.1)(600,89.1 )(800, 88.361)(1000,88.980)  };\addlegendentry{PEFS \cite{marahatta2020pefs}}
			\addplot[color=cyan, mark=triangle,thick] coordinates { 
				(0,0) (200,90.5)(400, 88.9)(600,89.7 )(800, 89.361)(1000,90.80)  };\addlegendentry{CDEF \cite{xu2018improving}}
			\addplot[color=orange, mark=o,thick] coordinates { 
				(0,0) (200,79.5)(400, 80.1)(600,81.1 )(800, 79.361)(1000,79.80)  };\addlegendentry{HDCC \cite{pinto2016hadoop}}
			\end{axis}
			\end{tikzpicture}}}\hfill

	\caption{Failure prediction accuracy for BB workload }
	\label{fig:accuracy_BB}
	
\end{figure}
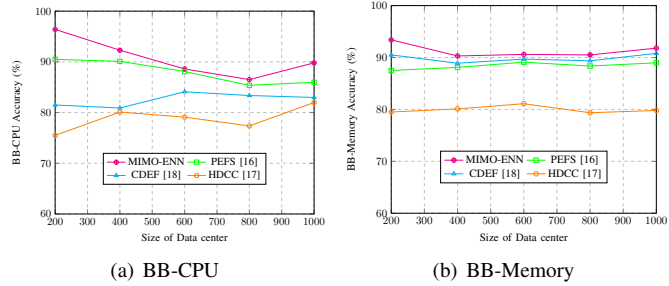

%\begin{figure}[!htbp]
	
%	\centering
	%\subfigure[GCD-CPU ]{\includegraphics[width=.192\textwidth]{Figures/gcdcpuaccuracy}}\hfill
	%\subfigure[GCD-Memory ]{\includegraphics[width=.192\textwidth]{Figures/gcdmem_accuracy}}\hfill
%	\subfigure[BB-CPU ]{\includegraphics[width=.241\textwidth]{Figures/bbcpuaccuracy}}
%	\hfill
%	\subfigure[BB-Memory ]{\includegraphics[width=.241\textwidth]{Figures/bbmemaccuracy}}
%	\hfill
	%\subfigure[BB-Bandwidth ]{\includegraphics[width=.22\textwidth]{Figures/bb_BWaccuracy}}
	
%	\caption{Failure prediction accuracy }
%	\label{fig:accuracy}
	
%\end{figure} 

\subsubsection{Failure Tolerance}
\begin{figure}[!htbp]
	
	\centering
	\subfigure[GCD  ]{\resizebox{0.24\textwidth}{!}{\begin{tikzpicture}
			\begin{axis}[
			width=0.55\textwidth,
			height=0.45\textwidth,
			ybar,
			ymin=0,
			ymax=400,
			bar width=8pt,
			legend style={at={(0.45,0.92)},
				anchor=north,draw=none,legend columns=1},
			ylabel={Avg. number of VM migration},
			xlabel={Size of Data center},
			symbolic x coords={200,400,600,800,1000},
			xtick=data,
			ymajorgrids=true,
			xmajorgrids=true,
			grid style=dashed,
			%nodes near coords,
			%nodes near coords align={vertical}
			]
			\addplot [color=magenta, fill=magenta!40!white!50!,pattern color=magenta, postaction={
				pattern=north east lines
			}] coordinates { 
				(200,12) (400,19) (600,23) (800,26) (1000,33) };\addlegendentry{FT-ERM }

			\addplot [color=black, fill=black!30!white!50!,pattern color=black!70!, postaction={
				pattern=north east lines
			}] coordinates { 
				(200,21) (400,36) (600,74) (800,123) (1000,159) };\addlegendentry{FT-ERM without S-Box}
			\addplot [color=blue, fill=blue!50!white!50!,pattern color=blue, postaction={
				pattern=north east lines
			}] coordinates { 
				(200,68) (400,136) (600,196) (800,235) (1000,342) };\addlegendentry{Without FT-ERM}

			\end{axis}
			\end{tikzpicture}}}\hfill
	\subfigure[Bitbrains ]{\resizebox{0.24\textwidth}{!}{\begin{tikzpicture}
			\begin{axis}[
			width=0.55\textwidth,
			height=0.45\textwidth,
			ybar,
			ymin=0,
			ymax=370,
			bar width=8pt,
			legend style={at={(0.45,0.9)},
				anchor=north,draw=none,legend columns=1},
			ylabel={Avg. number of VM migration},
			xlabel={Size of Data center},
			ymajorgrids=true,
			xmajorgrids=true,
			grid style=dashed,
			symbolic x coords={200,400,600,800,1000},
			xtick=data,
			%nodes near coords,
			%nodes near coords align={vertical}
			]
			\addplot [color=magenta, fill=magenta!40!white!50!,pattern color=magenta, postaction={
				pattern=north east lines
			}] coordinates { 
				(200,4) (400,6) (600,14) (800,13) (1000,18) };\addlegendentry{FT-ERM }

			\addplot [color=black, fill=black!30!white!50!,pattern color=black!70!, postaction={
				pattern=north east lines
			}] coordinates { 
				(200,19) (400,29) (600,44) (800,73) (1000,109) };\addlegendentry{FT-ERM without S-Box}
			\addplot [color=blue, fill=blue!50!white!50!,pattern color=blue, postaction={
				pattern=north east lines
			}] coordinates { 
				(200,76) (400,96) (600,105) (800,208) (1000,342) };\addlegendentry{Without FT-ERM}

			\end{axis}
			\end{tikzpicture}}}\hfill

	\caption{Average number of VM migration }
	\label{fig:VMmigration}
	
\end{figure}
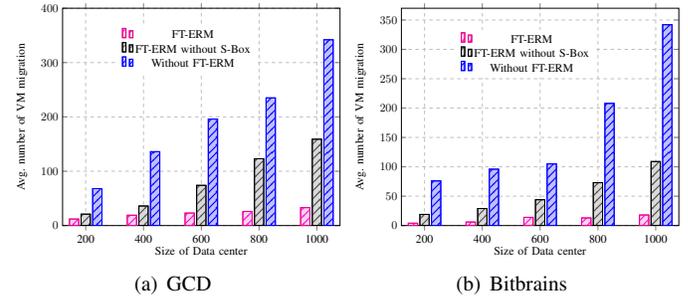

Fig. \ref{fig:VMmigration} compares the failure tolerance with respect to the average number of VM migrations which is least for FT-ERM, consecutively followed by FT-ERM without S-Box (i.e., with MIMO-ENN prediction) and without FT-ERM approaches. The achieved results depict that the number of VM migrations increases non-uniformly with the size of the datacenter. The average number of VM migration using FT-ERM varies from 3\% to 6\% of the size of the datacenter. However, FT-ERM significantly reduces live VM migrations up to 72.6\% and 88.6\% over FT-ERM without S-Box and without FT-ERM, respectively for GCD workload; and up to 79.9\% and 93.3\% against FT-ERM without S-Box and without FT-ERM, respectively for BB workload.
 
\subsubsection{Availability}
\begin{figure}[!htbp]
	\centering
	\includegraphics[width=0.75\linewidth]{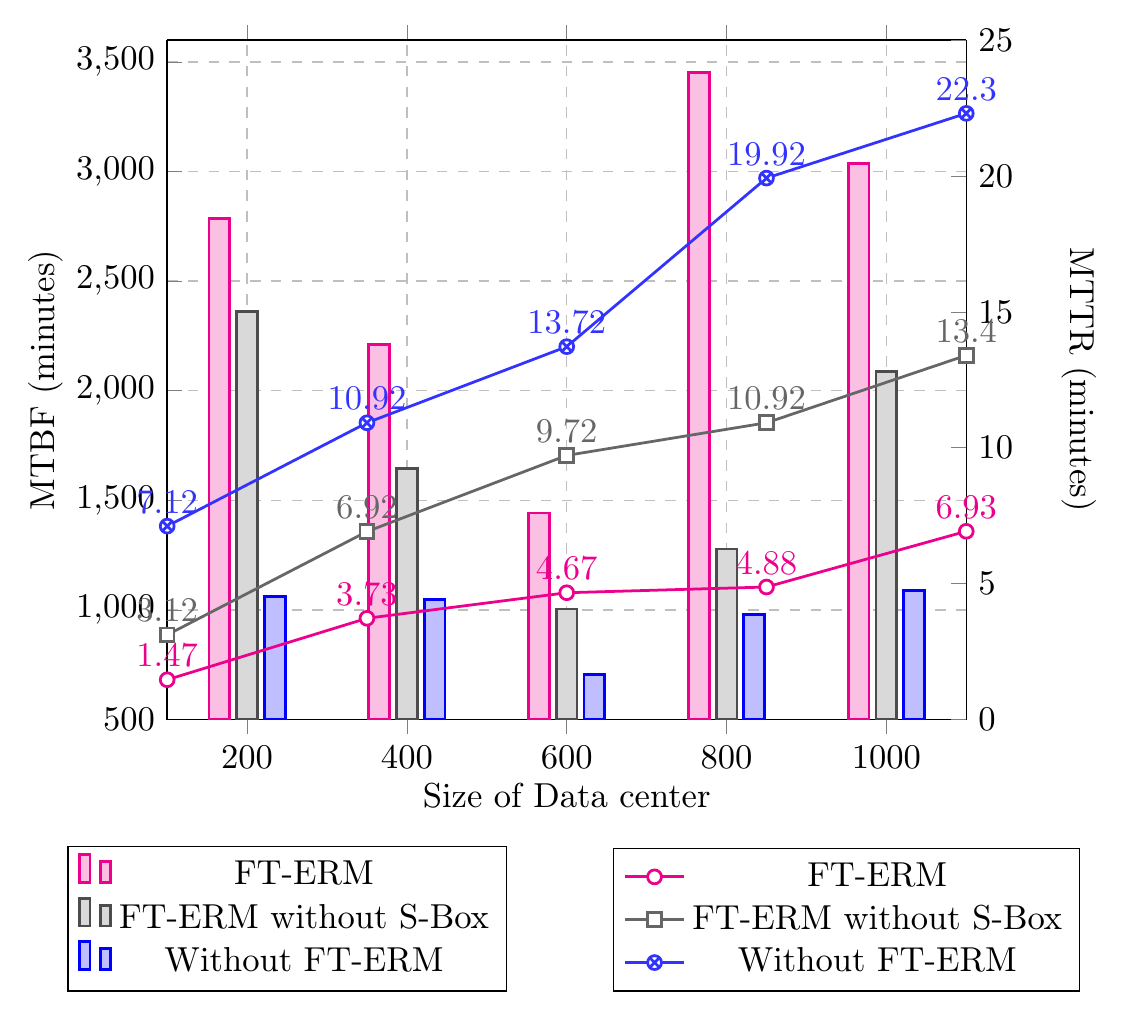}
	
	\caption{MTTR and MTBF of GCD workload}
	\label{fig:gcdmttrmtbf}
\end{figure}

\begin{figure}[!htbp]
	\centering
	
	\includegraphics[width=0.75\linewidth]{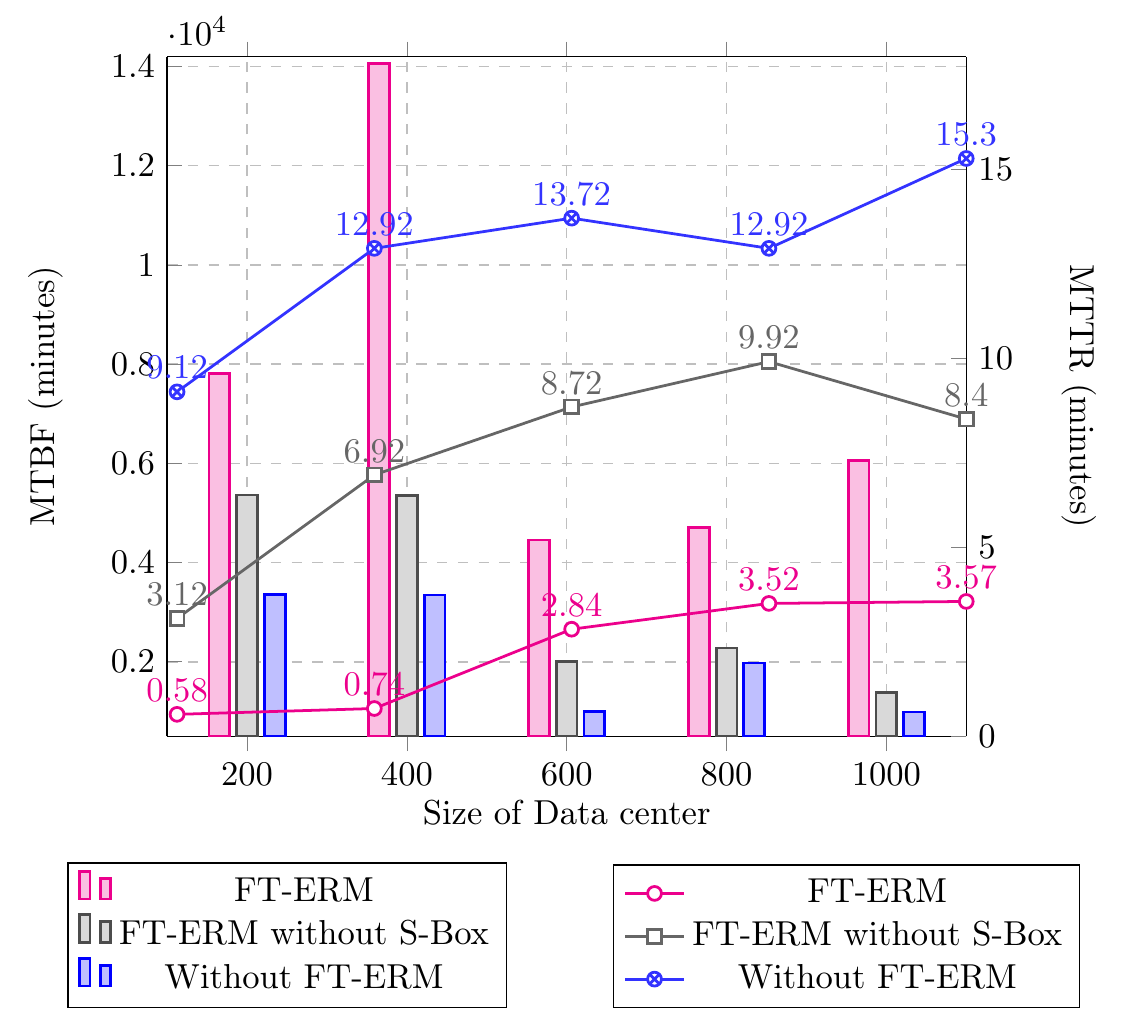}
	\caption{MTTR and MTBF of BB workload}
	\label{fig:bbmtbfmttr}
\end{figure}

\begin{figure}[!htbp]
	
	\centering
	\subfigure[GCD  ]{\resizebox{0.24\textwidth}{!}{
			\begin{tikzpicture}
			\begin{axis}[
			width=0.55\textwidth,
			height=0.45\textwidth,
			smooth,
			stack plots=y,
			xlabel={Size of Datacenter},
			ylabel={Availability ($\%$)},
			legend style={at={(0.5,-0.28)},
				anchor=north ,legend columns=1},
			area style,
			enlarge x limits=false]
			\addplot+[color=blue, fill=blue!50!white!50!,mark=otimes,thick] coordinates
			{ (1000,67.52) (800,66.5) (600, 64.52) (400, 65.8) (200, 64.12)}
			\closedcycle;\addlegendentry{Without FT-ERM}
			\addplot [color=black!70!,mark=square, fill=black!30!white!50!,thick]coordinates
			{ (1000,16.32) (800,18.5) (600, 22.59) (400, 21.02) (200, 17.27)}
			\closedcycle;\addlegendentry{FT-ERM without S-Box}
			
			\addplot [color=magenta, fill=magenta!50!white!50!,mark=o,thick] coordinates
			{(1000, 15.42) (800,14.2) (600, 11.4) (400, 12.52) (200, 17.2)}
			\closedcycle;\addlegendentry{FT-ERM}

			%\legend{GCD,PL, BB} 
			\end{axis}
			\end{tikzpicture}}}\hfill
	\subfigure[Bitbrains ]{\resizebox{0.24\textwidth}{!}{  \begin{tikzpicture}
			\begin{axis}[
			width=0.55\textwidth,
			height=0.45\textwidth,
			smooth,
			stack plots=y,
			xlabel={Size of Datacenter},
			ylabel={Availability ($\%$)},
			legend style={at={(0.5,-0.28)},
				anchor=north ,legend columns=1},
			area style,
			enlarge x limits=false]
			\addplot+[color=blue, fill=blue!50!white!50!,mark=otimes,thick] coordinates
			{ (1000,59.52) (800,58.5) (600, 66.52) (400, 57.8) (200, 65.12)}
			\closedcycle;\addlegendentry{Without FT-ERM}
			\addplot [color=black!70!,mark=square, fill=black!30!white!50!,thick]coordinates
			{ (1000,24.32) (800,26.5) (600, 21.59) (400, 27.02) (200, 22.27)}
			\closedcycle;\addlegendentry{FT-ERM without S-Box}
			
			\addplot [color=magenta, fill=magenta!50!white!50!,mark=o,thick] coordinates
			{(1000, 15.42) (800,14.2) (600, 11.4) (400, 14.52) (200, 12.2)}
			\closedcycle;\addlegendentry{FT-ERM}

			%\legend{GCD,PL, BB} 
			\end{axis}
			\end{tikzpicture}}}\hfill
	%\subfigure[Average Power consumption ]{\includegraphics[width=.32\textwidth]{Figures/Power}}

	\caption{Average Availability }
	\label{fig:availability}
	
\end{figure}
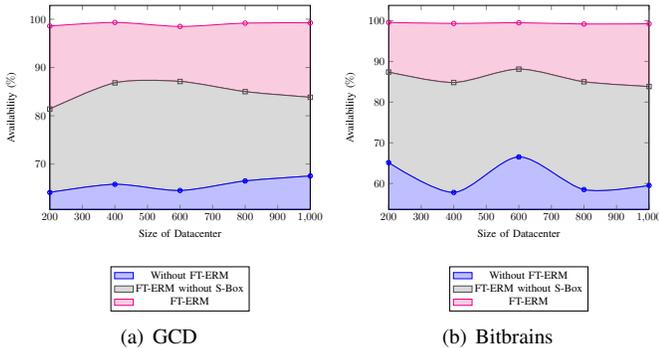

The availability varies with the values of MTBF and MTTR, obtained during online processing over time-interval \{$t_1$, $t_2$\}. The variations observed (during experimental simulation) in the values of MTBF and MTTR are shown in Figs. (\ref{fig:gcdmttrmtbf}) and (\ref{fig:bbmtbfmttr}) for GCD and BB workload execution, respectively. It is to be noticed that MTTR decreases when MTBF increases, which specifies an inverse relation between them.
{The MTBF values decreases while the MTTR increases with growing size of the datacenter because of the slight decrement in the percentage of accuracy during prediction of failures (Table \ref{table:performanceGCD} and Table \ref{table:performanceBB}).} The MTBF decreases and MTTR increases in the order: FT-ERM $<$ FT-ERM without S-Box $<$ FT-ERM without MIMO-ENN.
% On the same lines, MTTR increases such that FT-ERM $\leq$ OFP-RM without S-Box $\leq$ OFP-RM. 
 Fig. \ref{fig:availability} entails the availability for GCD (Fig. \ref{fig:availability}(a)) and BB (Fig. \ref{fig:availability}(b)) where FT-ERM outperforms FT-ERM without S-Box and without FT-ERM by 17.2\% and 34.47\%, respectively for GCD; and 11.4\% and 32.99\%, respectively for BB online workload distribution {due to precise estimation and mitigation of VM failures and lesser number of live VM migrations in case of FT-ERM as compared with rest of the two approaches.} 

\subsubsection{Load balancing metrics}
Table \ref{table:comparison_performanceGCD} and Table \ref{table:comparison_performanceBB} compare the power consumption ($PW$), resource utilization ($RU$), and temperature ($T$) of FT-ERM, FT-ERM without S-Box, and FT-ERM without MIMO-ENN frameworks for varying size of datacenters for GCD and BB workloads, respectively. The average power consumption using FT-ERM shows reduction of 38.83\% and 62.4\% for GCD and BB workloads, respectively against without FT-ERM framework.  
\begin{table}[!htbp]
	
	\caption[Table caption text] {Performance metrics for GCD workloads}  
	\label{table:comparison_performanceGCD}
	\small
	\centering 
	%\resizebox{0.8\textwidth}{!}{\begin{minipage}{\textwidth}
	\resizebox{8.5cm}{!}{
		\centering
		%\begin{tabular}{|p{1.2cm}|p{1.5cm}|p{0.6cm}|p{0.6cm}|p{0.6cm}|p{0.6cm}|p{0.6cm}|}
			\begin{tabular}{|l|l|c|c|c|c|c|c|}
			\hline
			
			Approach&Metrics&200 &400 &600&800&1000&Average \\ \hline \hline			
			\multirow{3}{*}{FT-ERM}	&$PW$(KW)& 7.72& 17.71& 27.05& 32.6&37.67& 24.55\\ \cline{2-8}
			&$RU$(\%)&60.46& 60.84&62.25 &61.05&63.03&61.15 \\ \cline{2-8}
			
			&$T$($^{\circ} C$)&22.4&21.5&22.2&22.4&22.8&22.26  \\ \hline 
	 \hline

		FT-ERM	&$PW$(KW)& 7.51&16.2&22.2&29.7&36.1& 22.34\\ \cline{2-8} %9.7&22.5&34.9&52.4&75.9  
		Without	&$RU$(\%)&59.50& 62.71& 62.31&61.60&62.20& 61.57 \\ \cline{2-8}
		
		{S-Box}	&$T$($^{\circ} C$)&23.2&23.4&22.6&23.6&23.5& 23.26 \\ \hline \hline	%\multirow{3}{*}
		Without	&$PW$(KW)& 13.32& 26.68&40.13&53.61&66.96&40.14   \\ \cline{2-8}
		\multirow{2}{*}{FT-ERM}	 &$RU$(\%)&58.20& 58.10&58.29&58.37&58.29&58.25\\ \cline{2-8}
			
	&$T$($^{\circ} C$)&25.30&25.20&25.20&25.31&25.30&25.26  \\ \hline

			\end{tabular}}
\end{table}
 The proposed FT-ERM framework improves the average resource utilization by 5.83\% and 5.97\% for GCD and BB workloads, respectively over without FT-ERM. However, FT-ERM consumes 9.9\% and 6.1\% more power than FT-ERM without S-Box for GCD and BB workloads, respectively. Also, it shows 0.42\% and 0.09\% lesser resource utilization over FT-ERM without S-Box for GCD and BB workloads, respectively. {This is due to allocation of VMs according to the size of S-Boxes determined by their respective clusters to avoid the risk of resource failures.} Moreover, FT-ERM maintains a balanced thermal condition in the datacenter by giving an average of 22.26$^{\circ} C$ which is lesser by 1$^{\circ} C$ and 2.66$^{\circ} C$ against FT-ERM without S-Box, and without FT-ERM, respectively for GCD workload. The temperature of datacenter is lower by 3.15$^{\circ} C$ and 3.69$^{\circ} C$ against FT-ERM without S-Box, and without FT-ERM frameworks, respectively for BB workload.
\begin{table}[!htbp]
	
	\caption[Table caption text] {Performance metrics for Bitbrains workloads}  
	\label{table:comparison_performanceBB}
	\small
	\centering 
	%\resizebox{0.8\textwidth}{!}{\begin{minipage}{\textwidth}
	\resizebox{8.5cm}{!}{
		\centering
		%\begin{tabular}{|p{1.2cm}|p{1.5cm}|p{0.6cm}|p{0.6cm}|p{0.6cm}|p{0.6cm}|p{0.6cm}|}
		\begin{tabular}{|l|l|c|c|c|c|c|c|}
			\hline
			
			Approach&Metrics&200 &400 &600&800&1000&Average \\ \hline \hline			
			\multirow{3}{*}{FT-ERM}	&$PW$(KW)&
			6.01 &11.00&15.91&21.02&26.03&16.00 \\ \cline{2-8}
			&$RU$(\%)&61.68& 62.23& 61.92&61.97&61.23&61.81 \\ \cline{2-8}
			
			&$T$($^{\circ} C$)&22.05&22.19&22.23&22.05&22.34&22.17

			\\ \hline \hline
			
			FT-ERM	&$PW$(KW)&5.02 &10.05&15.08&20.12&25.16&15.08
			 \\ \cline{2-8}  
			Without	&$RU$(\%)&61.40& 61.60& 61.50&62.60&62.40& 61.90 \\ \cline{2-8}
			
			{S-Box}	&$T$($^{\circ} C$)&25.30&25.33&25.32&25.32&25.33&25.32
			
			\\ \hline 
			\hline
			Without	&$PW$(KW)& 13.31&26.77&40.12&53.60&67.05&40.17   \\ \cline{2-8}
			\multirow{2}{*}{FT-ERM}	 &$RU$(\%)&58.16& 58.42&58.28&58.36&58.42&58.33\\ \cline{2-8}
			
			&$T$($^{\circ} C$)
			&25.00&25.70&26.50&25.60&26.50& 25.86  \\ \hline 
	 	
	\end{tabular}}
\end{table}

 \subsubsection{{FT-ERM v/s FAEE for Grid5000 Failure traces}}
 Table \ref{table:Grid5000comparison} compares the failure prediction accuracy, downtime and failure reduction percentage of the proposed FT-ERM framework and FAEE \cite{sharma2019failure} using Grid 500 Failure traces \cite{kondo2010failure}. These traces provide information about the failures and hardware configuration of approximately 1300 nodes. The MTBF and MTTR are calculated by using the failure start and stop time 
 information given in the traces. FT-ERM outperforms FAEE which is based on Exponential Smoothening based prediction approach by improving failure prediction accuracy by approximately 32.4\%; and by lowering the downtime and the failure reduction up to 41.3\% and 50.8\%, respectively. 
 \begin{table}[!htbp]
 	\centering
 	\caption[Table caption text] {FT-ERM v/s FAEE: Grid5000 FTA dataset }  %\cite[p.10]{refid} }
 	\label{table:Grid5000comparison}
 	%\resizebox{0.8\textwidth}{!}{\begin{minipage}{\textwidth}
 		\resizebox{8.5cm}{!}{
 	\begin{tabular}{|l|c|c|c|}
 		\hline			
 		Approach&Prediction accuracy& Avg. downtime & Number of failures \\ 	\hline
 		FAEE \cite{sharma2019failure} &57\%-71\%&approx. 15 Hrs. &4500 \\ 	
 		FT-ERM	&88\%-94\% &8.8 Hrs. &2215 \\ 	
 		
 		\hline	
 	\end{tabular}}
 	%\end{minipage}}
 \end{table}
 
\section{Conclusions and Future work}
%This work proposed a novel cloud resource management framework which predicts failure of VMs due to deficiency of one or multiple resources concurrently
A novel FT-ERM framework is proposed which embeds HA in VMs as well as servers for predicting any failure proactively and triggering necessary failure tolerance actions. An online MIMO-ENN failure predictor is developed for the prediction of multiple resource usage of VMs and  estimate their failure status in real-time. The framework employs a failure tolerance unit that decides safer allocation of failure prone VMs and migrates them to the selected server using clustering based S-Box mechanism. The performance evaluation shows that the proposed framework maximizes service availability and minimizes performance degradation due to overloads, cloud outages, SLA violations and excess power consumption. %All the results are supported by the experiments and comparison with state-of-art techniques which reveals that the proposed framework significantly improves the availability of cloud datacenter.  
{In the future, the proposed framework can be extended to achieve reactive fault tolerance by employing strategies such as N-Version Programming. Further, along with reliability, security can be included by considering trust among VMs before mapping to a selected server.}   
%\ifCLASSOPTIONcompsoc
  %\section*{Acknowledgments}
  %The author would like to thank National Institute of Technology, Kurukshetra, India for financially supporting the research work.
  
%\else
 % \section*{Acknowledgment}
%\fi

% authors would like to thank...
% \section*{Acknowledgments}
%The author would like to thank National Institute of Technology, Kurukshetra, India for financially supporting the research work.

\bibliographystyle{IEEEtran}
\bibliography{mybibfile}
\vskip 0pt plus -1fil 
\begin{IEEEbiography}[{\includegraphics[width=0.7\linewidth]{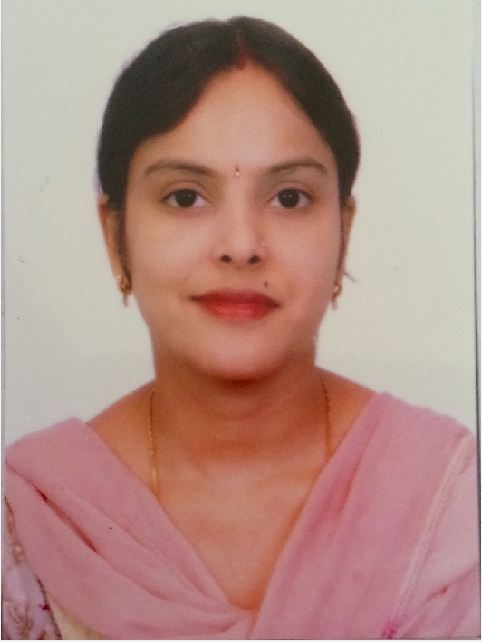}}]{Deepika Saxena}
	received her M.Tech degree in Computer Science and Engineering  from Kurukshetra University Kurukshetra, Haryana, India in 2014. Currently, she is pursuing her Ph.D from Department of Computer Applications, National Institute of Technology (NIT), Kurukshetra, India. Her major research interests are neural networks, predictive analytics, evolutionary algorithms, scheduling and security in cloud computing.	
\end{IEEEbiography}
\vskip 0pt plus -1fil 
\begin{IEEEbiography}[{\includegraphics[width=0.7\linewidth]{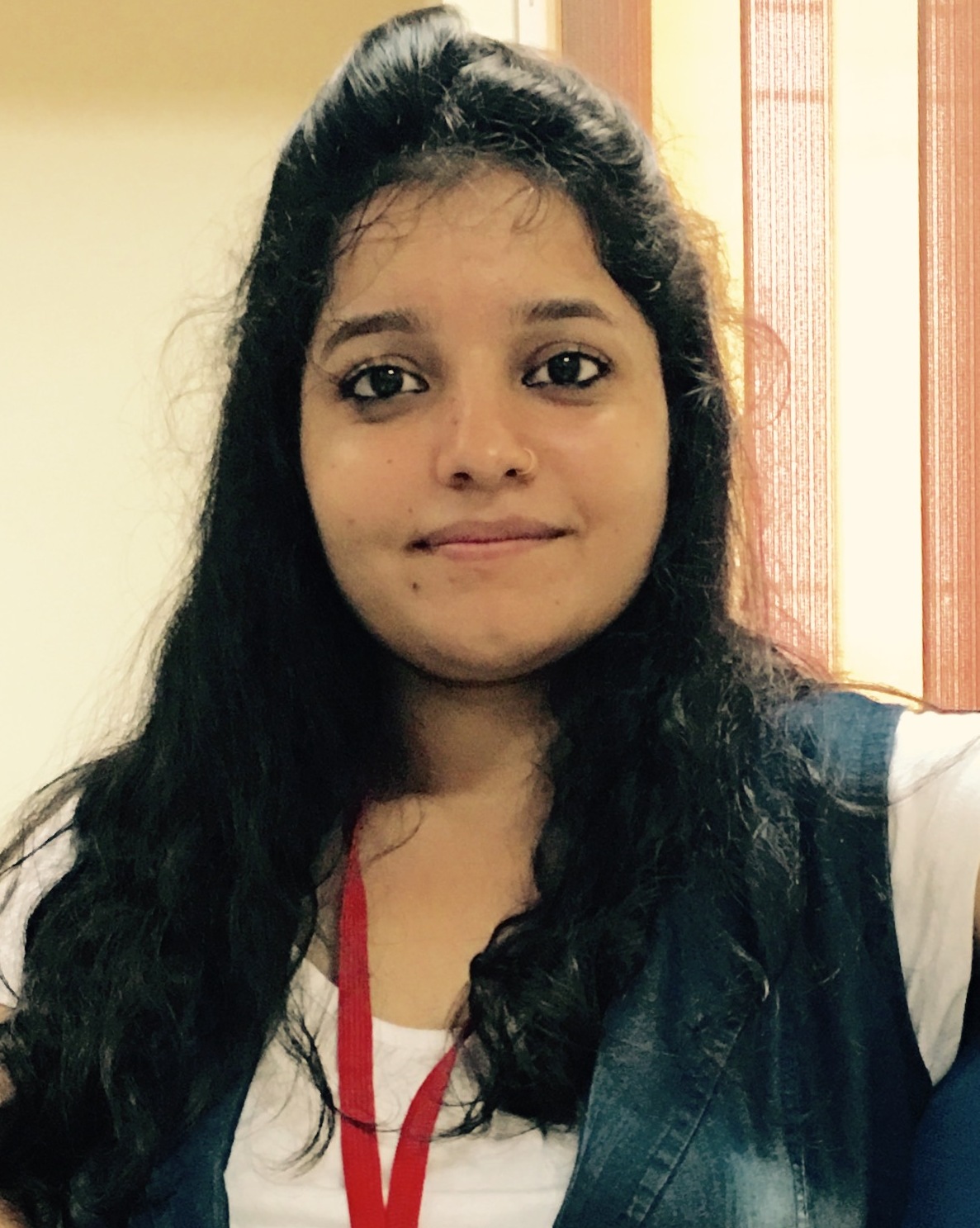}}]{Ishu Gupta} received BCA and MCA (Gold Medalist) degrees in Computer Science from Kurukshetra University, India, and Ph.D. from the Department of Computer Applications, National Institute of Technology (NIT), Kurukshetra, India. Currently, she is a post-doc fellow at National Sun Yat-Sen University, Kaohsiung, Taiwan. She is awarded, Senior Research Fellowship (SRF) by the University Grants Commission (UGC), Government of India. Her major research interests include the areas of Cloud Computing, IoT, Machine Learning, Information Security. 
	
\end{IEEEbiography}
\vskip 0pt plus -1fil 
\begin{IEEEbiography}[{\includegraphics[width=0.7\linewidth]{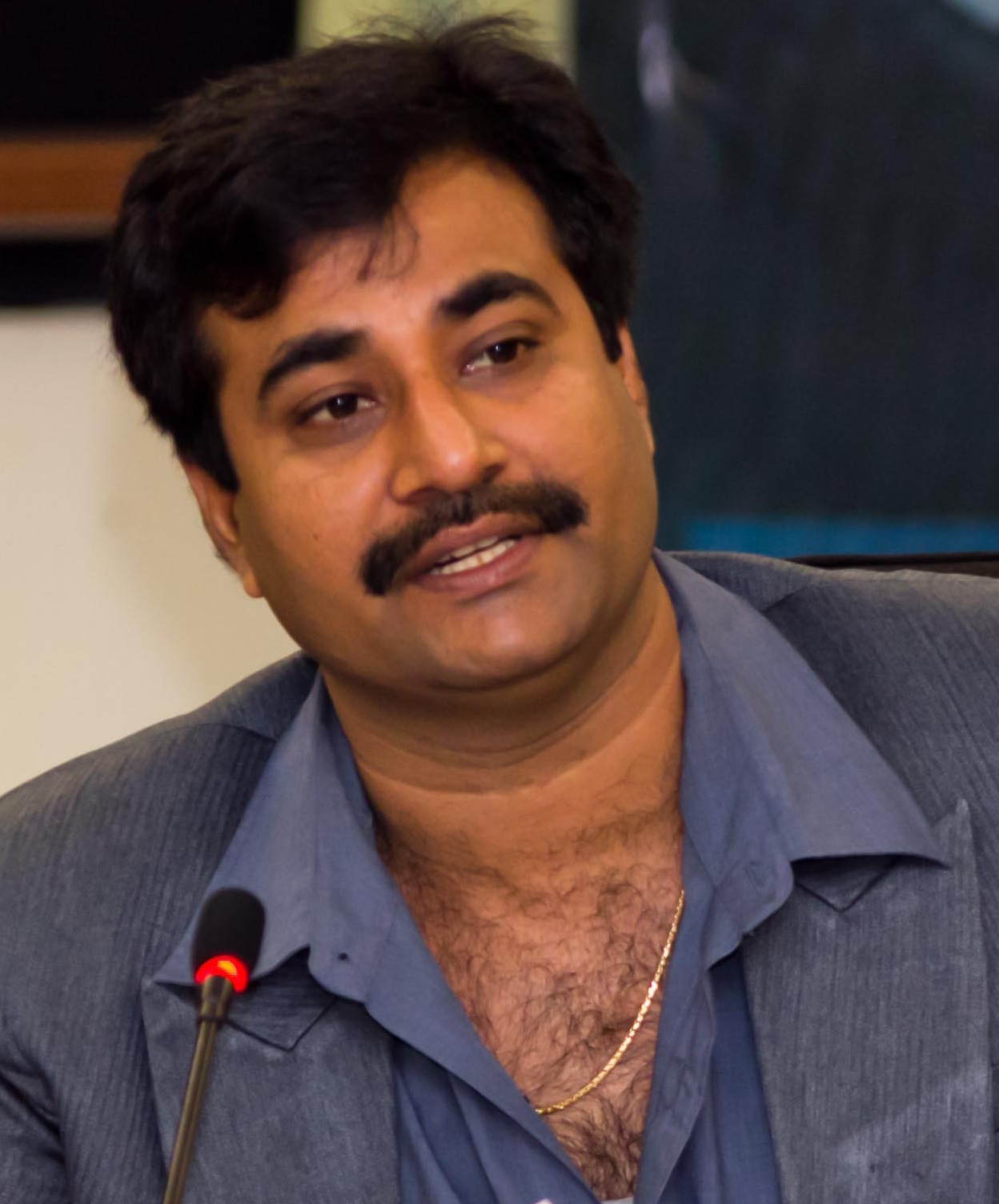}}]{Ashutosh Kumar Singh}
	is working as a Professor and Head in the Department of Computer Applications, National Institute of Technology Kurukshetra, India. He received his PhD in Electronics Engineering from Indian Institute of Technology, BHU, India and Post Doc from Department of Computer Science, University of Bristol, UK. His research area includes Design and Testing of Digital Circuits, Data Science, Cloud Computing, Machine Learning, Security. He has published more than 300 research papers in different journals of high repute.	
\end{IEEEbiography}
\vskip 0pt plus -1fil 
\begin{IEEEbiography}[{\includegraphics[width=0.7\linewidth]{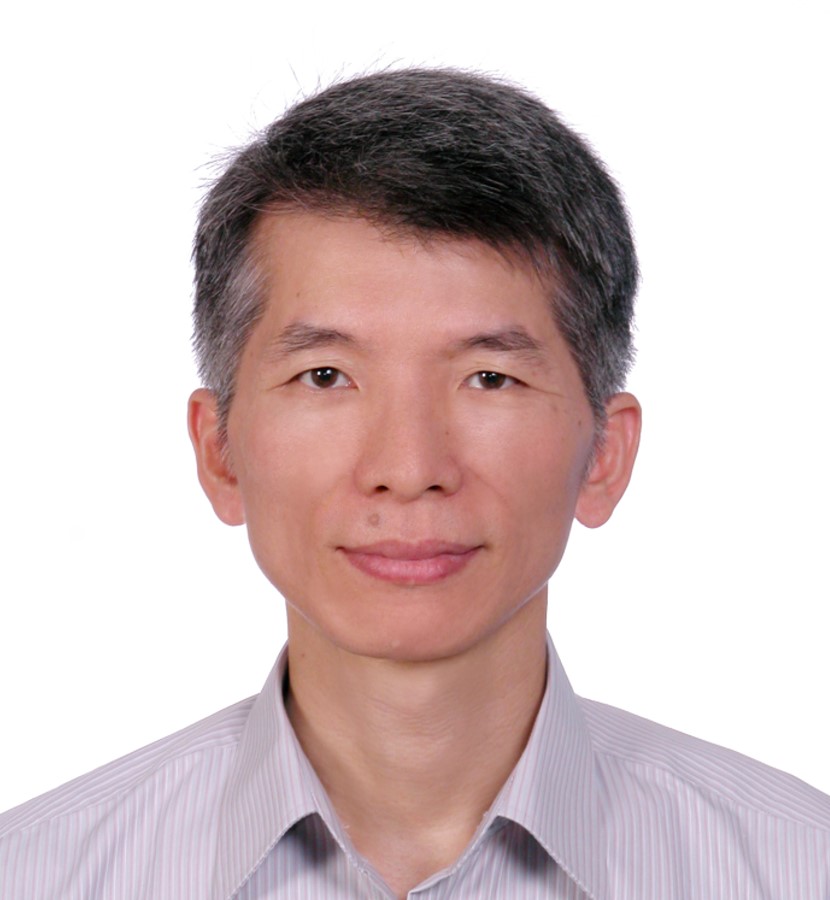}}]{Chung-Nan Lee}
	received B.S. and M.S. degrees, both in electrical engineering, from National Cheng Kung University, Tainan, Taiwan, in 1980 and 1982, respectively, and the Ph.D. degree in electrical engineering from the University of Washington, Seattle, in 1992. Since 1992, he has been with National Sun Yat-Sen University, Kaohsiung, Taiwan. Currently, he is a Distinguished Professor and Director of Cloud Computing Research Center. He was President of Taiwan Association of Cloud Computing from 2015 to 2017; VP for TA of APSIPA from 2019-2020. His current research interests include multimedia over wireless networks, cloud computing, and IOT.
\end{IEEEbiography}

\end{document}